\documentclass[onecolumn,appendixfloats]{emulateapj}  

\usepackage{color}
\usepackage{amsmath}

\def\simge{\mathrel{%
    \rlap{\raise 0.511ex \hbox{$>$}}{\lower 0.511ex \hbox{$\sim$}}}}
\def\simle{\mathrel{
    \rlap{\raise 0.511ex \hbox{$<$}}{\lower 0.511ex \hbox{$\sim$}}}}

\slugcomment{Published in {\it The Astrophysical Journal} vol.\ 843:122 (23pp), 2017 July 10}
\shorttitle{Cosmic Shoreline 2016}
\shortauthors{Zahnle \& Catling}

\begin{document}



 \title{The cosmic shoreline: the evidence that escape determines which planets have atmospheres, and what this may mean for Proxima Centauri b}  













\author{Kevin J. Zahnle\altaffilmark{1}} 
\affil{NASA Ames Research Center, Moffett Field, CA 94035}

\and

\author{David C. Catling\altaffilmark{2}}
\affil{Department of Earth and Space Sciences/Astrobiology Program, University of Washington, Seattle, WA}

\altaffiltext{1}{kevin.j.zahnle@NASA.gov}
\altaffiltext{2}{dcatling@u.washington.edu}


\begin{abstract}

The planets of the Solar System divide neatly between those with atmospheres and those without when 
arranged by insolation ($I$) and escape velocity ($v_{\mathrm{esc}}$).
The dividing line goes as $I \propto v_{\mathrm{esc}}^4$.
Exoplanets with reported masses and radii are shown to crowd against the extrapolation of the Solar System trend,
making a metaphorical cosmic shoreline that unites all the planets.  
The $I \propto v_{\mathrm{esc}}^4$ relation may implicate thermal escape.
We therefore address the general behavior of hydrodynamic thermal escape models ranging from Pluto to highly-irradiated Extrasolar Giant Planets (EGPs).  
 Energy-limited escape is harder to test because copious XUV radiation is mostly a feature of young stars,
 and hence requires extrapolating to historic XUV fluences ($I_{\mathrm{xuv}}$) using proxies and power laws.
 An energy-limited shoreline should scale as  
$I_{\mathrm{xuv}} \propto v_{\mathrm{esc}}^3\sqrt{\rho}$, which differs distinctly from the apparent $I_{\mathrm{xuv}} \propto v_{\mathrm{esc}}^4$ relation.     
 Energy-limited escape does provide good quantitative agreement to the highly irradiated EGPs.
 Diffusion-limited escape implies that no planet can lose more than 1\% of its mass as H$_2$.    
 Impact erosion, to the extent that impact velocities $v_{\mathrm{imp}}$ can be estimated for exoplanets,
 fits to a $v_{\mathrm{imp}} \approx 4\,-\,5\, v_{\mathrm{esc}}$ shoreline.
 The proportionality constant is consistent with what the collision of comet 
 Shoemaker-Levy 9 showed us we should expect of modest impacts in deep atmospheres.
 With respect to the shoreline, Proxima Centauri b is on the metaphorical beach.
 Known hazards include its rapid energetic accretion, high impact velocities, its early life on the wrong side of the runaway greenhouse,
 and Proxima Centauri's XUV radiation.  In its favor is a vast phase space of unknown unknowns.  
 
\end{abstract}



\section{Introduction}

Volatile escape is the classic existential problem of planetary atmospheres \citep{Hunten1990}.
The problem has gained new currency with the discovery and characterization of exoplanets \citep{Borucki2010,Lissauer2014}.
When escape is important it is likely to be rapid, and therefore relatively unlikely to be caught in the act,
but the cumulative effects of escape should show up in the statistics of the new planets \citep{Owen2013,Lopez2014}.
In this paper we discuss the empirical division between planets with and without apparent atmospheres inside and outside of the Solar System. 
The paper is organized around four figures that compare the planets (dwarf and other) of the
Solar System to the $\sim 590$ exoplanets that were relatively well-characterized as of 26 August 2016.
Then in Section \ref{section:five} we address the place of Proxima Centauri b and Trappist 1f among these planets.  

In Section 2 we compare total stellar radiation intercepted by a planet (insolation, $I$) to the planet's escape velocity ($v_{\mathrm{esc}}$).
In previous work we found that on such a plot the empirical division between planets with and without atmospheres
follows a $I\propto v^4_{\mathrm{esc}}$ power law that we have called the ``cosmic shoreline'' \citep{Zahnle1998a,Catling2009,Zahnle2013}.
We then compare the planets to the predictions of two different thermal escape models, one pertinent to small planets
with condensed volatiles at the surface, and the other pertinent to giant planets that are close to their stars.
We also compare the planets to the water vapor runaway greenhouse limit, which has a different relation to insolation and gravity.
    
In Section 3 we restrict stellar irradiation to extreme ultraviolet (EUV) and X-rays,
here lumped together as XUV radiation ($I_{\mathrm{xuv}}$).  
The intent is to address the popular XUV-driven escape hypothesis
\citep[e.g.,][the list is not complete]{Hayashi1979,Sekiya1980a,Sekiya1981,Watson1981,Lammer2003,Lecavelier2004,Yelle2004,
Tian2005,Erkaev2007,Garcia2007,Koskinen2009,Lammer2009,Murray-Clay2009,Tian2009,Erkaev2013,Koskinen2013,Lammer2013,
Owen2013,Koskinen2014,Lammer2014,Lopez2014,Tian2015,Erkaev2015,Erkaev2016,Owen2016}.
We estimate historic XUV fluences of the host stars to find
that the shoreline also appears in XUV with the same power law, $I_{\mathrm{xuv}}\propto v^4_{\mathrm{esc}}$. 

Section 4 begins by pointing out that the empirical $I_{\mathrm{xuv}}\propto v^4_{\mathrm{esc}}$ relation 
is not in accord with the basic predictions of energy-limited escape.
We compare the predictions of the simplest quantitative energy-limited escape model to the apparent planets. 
Because XUV radiation is well-suited to driving the escape of hydrogen in particular,
it is natural to discuss selective escape and diffusion-limited escape in the context of XUV-driven escape.
We show here that diffusion-limited escape, where applicable, reduces to a general result 
that may be particularly germane to super-earths.  

Section 5 addresses impact erosion.
Impact erosion of planetary atmospheres offers a plausible alternative to thermal or irradiation-driven escape \citep{Zahnle1992,Zahnle1998a,Catling2013,Schlichting2015}.
Here we compare impact velocities $v_{\mathrm{imp}}$ to escape velocities for the 
planets plotted in the previous figures.
 With impact erosion there is a reasonable expectation that a $v_{\mathrm{imp}} \propto v_{\mathrm{esc}}$ 
 relationship should apply at all scales; the difficulty in testing the hypothesis is in estimating what the 
 impact velocities should be.
 
 Section 6 addresses Proxima Centauri b (which we will refer to as ``Proxima b'' for the rest of the paper because 
 the star, which is uniquely the Sun's nearest neighbor, should really be just ``Proxima'').
In Section 6 we document how we plot Proxima b against the
 other planets and discuss its place with respect to the cosmic shoreline.

\section{Total insolation vs.\ escape velocity} 
\label{section:two}

\begin{figure}[!htb] 
   \centering
   \includegraphics[width=1.0\textwidth]{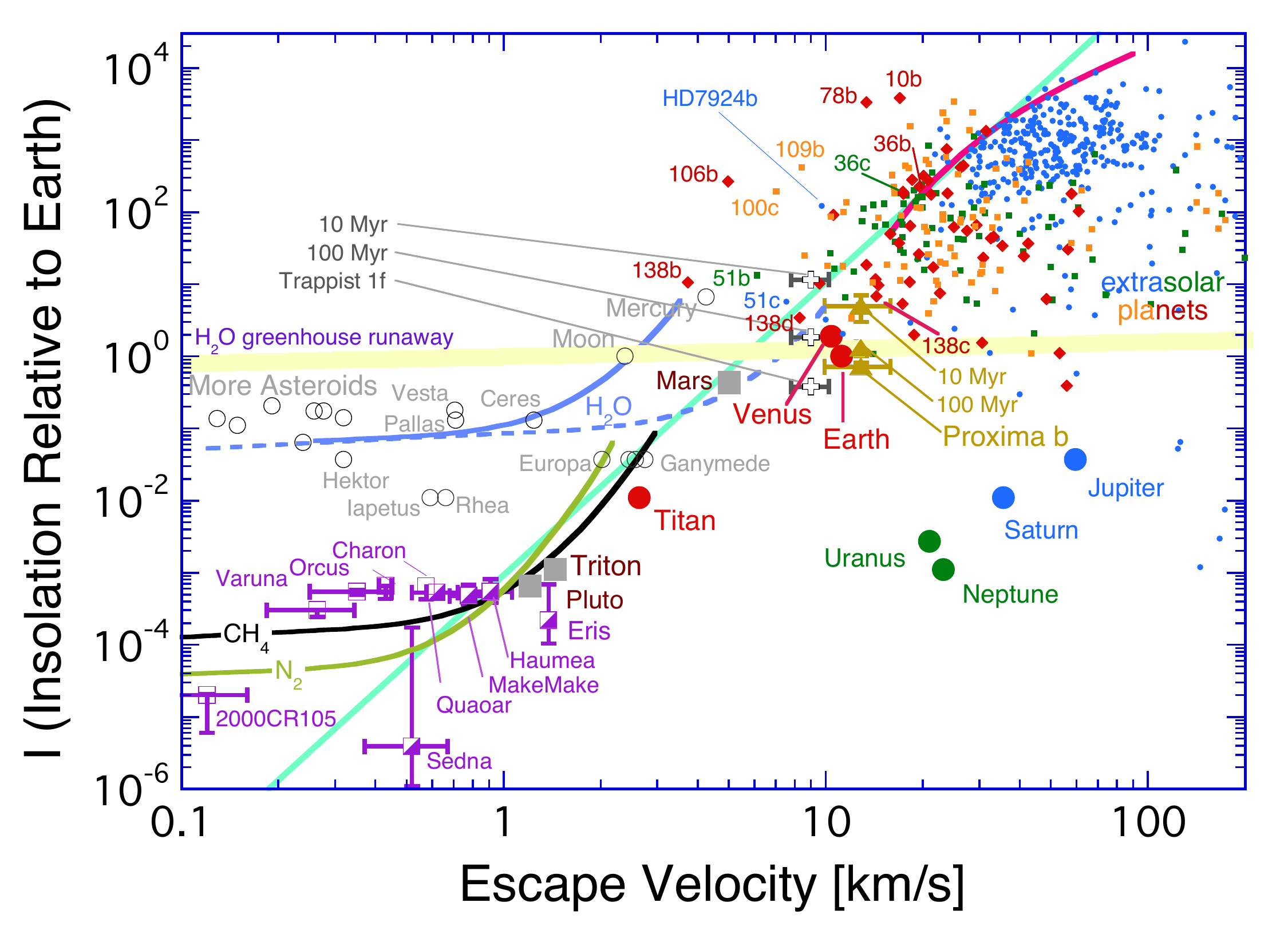} 
   \caption{\small  Atmospheres are found where gravity 
    --- here represented by the escape velocity --- is high and insolation 
   --- here represented by the total stellar insolation at the planet relative to that received by Earth --- is low.
  The presence or absence of an atmosphere on solar system objects is indicated by filled or open symbols, respectively. 
   Extrasolar planets with known masses and radii as of 26 August, 2016 \citep[][http://exoplanet.eu]{Schneider2011} are also plotted. 
   The extrasolar planets are presented as blue disks if Saturn-like ($R>8 R_{\oplus}$),
as green boxes if Neptune-like ($8 R_{\oplus} > R > 3 R_{\oplus}$),
 as red diamonds if Venus-like ($ R < 1.6 R_{\oplus}$),
 and as orange squares if none of the above ($3 R_{\oplus} > R > 1.6 R_{\oplus}$). 
  The simple $I \propto v_{\mathrm{esc}}^4$ power law bounding the cosmic shoreline is drawn by eye.
  Several of the more outlandish worlds are labeled by name; abbreviations like ``138b'' refer to ``Kepler 138b.'' 
  Error bars on the extrasolar planets are omitted for clarity. Planets that plot to the left of the
  $I \propto v_{\mathrm{esc}}^4$ power law may have very uncertain $v_{\mathrm{esc}}$ or they may be airless.
   Several models are also shown for comparison.
   Hydrodynamic thermal escape models for CH$_4$, N$_2$, and H$_2$O assume vapor pressure equilibrium at the surface.
   The two H$_2$O curves are for escape as H$_2$O (solid) or as H$_2$ (dashed).
   The magenta line is the thermal stability limit for hot Extrasolar Giant Planets (EGPs)  as described in Section \ref{section:EGP}.
   The curvature is caused by tidal stripping. 
   Also shown is the runaway greenhouse threshold for steam atmospheres.
   The models are described in the text.
   Proxima b is plotted in gold at three times (10 Myr, 100 Myr, now) in the luminosity evolution of Proxima.
   Trappist 1f  is plotted at three times (10 Myr, 100 Myr, now). 
 }
\label{Insolation}
\end{figure}

Figure \ref{Insolation} plots relative insolation $I$ against escape velocity $v_{\mathrm{esc}}$ for the adequately characterized planets.
The figure shows that atmospheres are found where gravitational binding energy is high and total insolation is low.
The figure also shows that the boundary between planets with and without apparent atmospheres
 is both well-defined and follows an empirical 
$I\propto v^4_{\mathrm{esc}}$ power law that extends from the planets of our solar system up through the known exoplanets.  

For most solar system objects we use escape velocities from \citet{Zahnle2003}.
The Kuiper Belt Objects (KBOs) and Pluto/Charon have been updated to reflect more recent information
\citep{Rabinowitz2006,Brown2007,Brown2013,Sicardy2011,Pal2012,Braga2013,Fornasier2013,Lellouch2013,Wesley2013,Stern2015}.
The presence or absence of an atmosphere for solar system objects is indicated by filled or open symbols, respectively.
 For the solar system, what is meant by ``having an atmosphere'' is usually pretty obvious,
 but even here there are borderline cases,
 such as Io, which has a very thin volcanogenic SO$_2$ atmosphere.
The KBOs are more ambiguous.
A few of the largest retain stores of frozen methane on their surfaces.
These are plotted in purple on Figure \ref{Insolation} as boxes-half-full.
It is reasonable to suppose that their surface volatiles will evaporate when close to the Sun and
they will then for a time have atmospheres similar to those of Pluto and Triton.
This sort of seasonal transformation is very likely for Eris, which is currently near aphelion,
and rather unlikely for Sedna, which is rather near its perihelion.

Figure \ref{Insolation} includes the nearly complete roster of transiting exoplanets with published radii and masses as of 26 August, 2016.
For these planets, orbital parameters, diameters, and stellar luminosities are measured, 
albeit not always very precisely.
 Apart from correcting Kepler 138c and 138d, 
  we have not attempted to weight the catalog by the purported quality of the data.
A few planets have been omitted because we were unable to estimate insolation.
The exoplanet.eu database \citep[][http://exoplanet.eu]{Schneider2011} was filtered to include only exoplanets with (i) a reported radius $R$; (ii) a reported mass $M$; and (iii) a reported orbital period $P$.
Radii and masses were used to compute escape velocities $v_{\mathrm{esc}}^2 = 2GM/R$.
For most of the exoplanets the measured diameters and densities are typical of giant planets,
which indicates that most of the transiting planets plotted on Figure \ref{Insolation} have atmospheres.
Planets with radii $R_p > 8 R_{\oplus}$ --- Saturns and super-Saturns --- are plotted as blue disks.
Planets with radii $1.6 R_{\oplus} < R_p < 8 R_{\oplus}$ --- Neptunes and sub-Saturns --- are plotted as dark blue disks.
Exoplanets with radii $R_p < 1.6 R_{\oplus}$ --- sub-Neptunes, Earths, and super-Earths --- are plotted as green diamonds.
This particular sorting by size appears to have some basis in fact \citep{Lopez2014}, with the three categories 
loosely corresponding to H$_2$-rich planets, H$_2$O-rich planets or H$_2$-enveloped rocky planets, and rocky planets.  

  For stellar luminosity we use the stellar radii $R_{\star}$ and stellar effective temperatures $T_{\star}$
  listed for most of the central stars in the exoplanet.eu database.
  \begin{equation} \label{L1star}
  L_{1\star} = 4\pi R_{\star}^2 \sigma T_{\star}^4 .
  \end{equation}
 If one or both of $R_{\star}$ and $T_{\star}$ is not listed,
 we estimate stellar luminosity using the visual magnitude $m_v$ and stellar distance $d$ (in parsecs) as reported in the exoplanet.eu database.
 \begin{equation} \label{L2star}
  L_{2\star} =  L_{\odot}\left(d/10\right)^2\,100^{0.2\left(4.83-m_v\right)}.
 \end{equation}
 Here $m_v=4.83$ is the visual magnitude of the Sun at 10 pc. 
 For a few of the planets neither $L_{1\star}$ nor $L_{2\star}$ can be computed; these planets we omit.
  Insolation $I$ is plotted as a dimensionless ratio relative to what is incident on Earth today.
 Insolation is computed using the planet's reported semimajor axis $a$ if available,
 otherwise we estimate $a$ from the stellar mass $M_{\star}$ and the period $P$ 
 \begin{equation} \label{Keplers_Law}
  a^3 = \frac{G M_{\star} P^2 }{4\pi^2}
 \end{equation}
 in order to compute 
 \begin{equation} \label{I_vs_Earth}
 I =\frac {L_{\star}}{L_{\odot}} \frac{a_{\oplus}^2}{a^2} 
 \end{equation}
 where $a_{\oplus}$ is the semimajor axis of Earth's orbit. 

The general pattern seen in Figure \ref{Insolation} is what one would expect to see if escape were the most important
process governing the volatile inventories of planets.
Where the gravitational well is deep (measured by escape velocity),
or where the influence of the central star is weak (measured here by insolation), planetary atmospheres are thick.
Where the gravity is weak or the star too bright, there are only airless planets.
It is curious that the non-giant planets in our Solar System that have atmospheres seem
to be strung out along a single $I\propto v_{\mathrm{esc}}^4$ line,
rather than scattered over the half-space below and to the right of the bounding line.
A second surprise is that the known transiting extrasolar giant planets (EGPs) crowd against the extrapolation of that line.
 The $I\propto v_{\mathrm{esc}}^4$ line spans two orders of magnitude in escape velocity and nearly eight orders
 of magnitude in insolation.
 It is remarkable that a single power law should appear to relate hot EGPs at one extreme to Pluto and Triton at the other.

What Figure \ref{Insolation} does not show is what one might expect to see if the presence
of atmospheres depended mostly on local nebular temperature (during accretion) or volatile supply.
To put this another way, small warm worlds with significant atmospheres are missing.
Such worlds are permitted if not mandatory in supply-side scenarios;
indeed, active comets provide extreme examples of what such worlds can look like during their brief lives.

\subsection{Thermal escape from icy planets}
\label{section:CC}

We consider isothermal atmospheres in which the major gas is in vapor pressure equilibrium with condensed volatiles at the surface.
We call these Clausius-Clapeyron (CC) atmospheres because they are controlled by vapor pressure \citep{Lehmer2017}.
Clausius-Clapeyron atmospheres are fairly common in the solar system, with Pluto, Triton, and Mars being
 notable examples.  
Our model resembles an earlier Jeans escape model used by \citet{Schaller2007} and \citet{Brown2012} to address thermal escape from Kuiper Belt objects.
The chief difference is that we are addressing much faster escape rates, 
conditions where Jeans escape would be misapplied.
\citet{Perez-Becker2013} investigated similar models for evaporation of very hot silicate planets.
Limitations and uncertainties stemming from the isothermal approximation are discussed along
with some alternative approximations to planetary winds that may seem more realistic are discussed in section \ref{discussion}, 
 and examined in more detail in the Appendices.

Isothermal hydrodynamic escape is described in terms of the isothermal sound speed 
$c_{\circ}^2 = k_B T_c/ m$, where $T_c$ is the temperature and $m$ is the mean molecular mass of the atmosphere.
Pressure $p$ is that of a perfect gas $p=\rho\, c_{\circ}^2 $ with density $\rho$.
The outflow is described in spherical (radial) geometry in steady state by continuity
\begin{equation}
\label{continuity}
\frac{\partial }{\partial r}\left(\rho u r^2 \right) = 0, 
\end{equation}
where $u$ is the outflow velocity and $r$ is the radial coordinate increasing outward,
and by the force balance of inertia, pressure, and gravity, 
\begin{equation}
\label{force}
{u}\frac{\partial u }{ \partial r} + \frac{1}{ \rho} \frac{\partial p}{ \partial r} =  -\frac{GM }{ r^2} ,
\end{equation}
in which $M$ is the mass of the planet and $G$ is the universal gravitational constant. 
These can be combined into a single isothermal planetary wind equation, 
\begin{equation}
\label{parker}
\left(u^2 - c_{\circ}^2\right) \frac{1}{ u} \frac{\partial u }{ \partial r} = \frac{2c_{\circ}^2}{ r} - \frac{GM }{ r^2} .
\end{equation}
In a strongly bound atmosphere, $u^2 \ll c_{\circ}^2$ and $2c_{\circ}^2 \ll GM/r^2$ near the surface, so that $\partial u/\partial r >0$.
At large radii the geometric term $2c_{\circ}^2/r$ term eventually surpasses the gravity term and the right hand side of Eq \ref{parker} changes sign. 
At the critical distance $r_c$ where $2c_{\circ}^2/r=GM/r^2$,
zeroing the left hand side of Eq \ref{parker} requires either that $u^2=c_{\circ}^2$ or that $\partial u/\partial r=0$.
The unique solution with $\partial u/\partial r >0$, in which the velocity is equal to the sound speed
$u_c=c_{\circ}$ at the critical point $r_c=GM/2c_{\circ}^2$, is the critical transonic solution for the wind and is the physical
solution provided that the ram pressure of the wind $\rho u_c^2$ at the critical point 
exceeds any background pressure exerted by interplanetary space. 

Equation \ref{parker} is easily integrated for the upward velocity at the surface, $u_s$,
in terms of the isothermal temperature $T_c$ and the critical point conditions.   
Where escape is modest and $u_s^2 \ll c_{\circ}^2$,  
\begin{equation}
 u_s \approx c_{\circ} \left( \frac{r_c}{ r_s}\right)^2 \exp{\left( \frac{3}{2} - \frac{GM}{c_{\circ}^2 r_s} \right)}
 \end{equation}
gives a good approximation to $u_s$.
The density and pressure at the surface of a CC atmosphere are determined by the saturation vapor pressure at the surface temperature $T_s$,
\begin{equation}
  \rho_s = \frac{m \, p_s(T_s) }{ k_B T_s}  .
\end{equation}
For display in Figure 1 we use empirical expressions for the vapor pressures of CH$_4$, N$_2$, and H$_2$O given by \citet{Fray2009}.
For the isothermal CC atmosphere the rate of mass loss is a function of $T_s$ and $u_s$
\begin{equation}
  {\dot M} = 4\pi\rho_s u_s r_s^2 .
  \end{equation}
  The other determining equation is the balance between the power absorbed from sunlight and the sum of power radiated in the thermal infrared and power spent
  evaporating, heating, and lifting warm gas to space,  
\begin{equation}
\label{energy_balance}
 \frac {\pi R^2 L_{\star}}{ 4 \pi a^2} = \frac{4\pi R^2 \sigma T^4_s}{ 1-\alpha} + \frac{GM{\dot M}}{ R} 
 + {\dot M}L_v,
\end{equation}
where $\alpha$ is the Bond albedo, which for icy worlds we will take as $\alpha=0.4$.
The rightmost term takes into account the latent heat of vaporization $L_v$, an important part of the energy budget for small icy bodies.   
Here we are interested in whether a planet can hold an atmosphere for billions of years.
In these cases escape is slow enough that the terms in Eq \ref{energy_balance} 
involving ${\dot M}$ are negligible,
so that the surface temperature is just the effective temperature.

\begin{table}[htp]
\caption{Surface temperatures and mesosphere temperatures of CH$_4$-rich atmospheres}
\begin{center}
\begin{tabular}{lrrrl}
Planet  &  Surface $T_s$ & Mesosphere $T_c$ &  $T_c/T_s$ & Reference\\
\hline
Titan   &  94$\phantom{^{\ast}}$   &  150-190  &  0.50 - 0.63 & \citet{Koskinen2011}\\
Triton  &  38$\phantom{^{\ast}}$    &   50-60    &  0.63 - 0.76 & \citet{Olkin1997} \\
Pluto  &  42$\phantom{^{\ast}}$    &   72    &  0.6 & \citet{Gladstone2016} \\
Uranus  &  60$^{\ast}$   &  110-150 &  0.4 - 0.55 & \citet{Marten2005}\\
Neptune  &  60$^{\ast}$   &  110-150 &  0.4 - 0.55 & \citet{Marten2005} \\
\hline
\multicolumn{4}{l}{$^{\ast}$ These are the effective temperatures.}
\end{tabular}
\end{center}
\label{table_one}
\end{table}
Empirically, upper atmospheres of the CH$_4$-rich planets in the solar system are roughly twice as hot
at high altitudes as they are at the surface (Table \ref{table_one}).
The upper atmospheres are warm because (i) CH$_4$ absorbs sunlight and (ii) CH$_4$ photolysis produces organic molecules and hazes that
absorb sunlight, but at low temperatures neither CH$_4$ nor the hazes radiate as effectively as they absorb. 
The upper atmosphere temperatures are higher on Uranus and Neptune because the background
gas is H$_2$, so that additional radiative coolants are limited to hydrocarbons like C$_2$H$_2$.
On Titan, Triton, and Pluto, the background gas is N$_2$, which enables production of
a wider variety of more efficient radiative coolants, HCN especially.
For the CH$_4$-N$_2$ atmospheres of icy worlds we let reality be a guide and
approximate the atmospheric temperature with $T_s = 0.6 \,T_c$. 

The black and green curves labeled ``CH$_4$'' and ``N$_2$'' are evaporation lines for the small icy planets.
The density is set to $\rho=2$ g/cm$^3$, similar to the densities of Triton, Titan, Ganymede, Callisto, and Pluto.
The curves are computed for a star of age $\tau_{\star}= 5$ billion years and an atmophile mass fraction $\Delta M/M= 0.01$,
where $\Delta M = \dot{M}\tau_{\star}$.
It is interesting that the nominal curves ``CH$_4$'' and ``N$_2$'' resemble what is actually seen in the solar system.
However, the results shown here are quite sensitive to $T_c/m$ (the sound speed, squared), 
and consequently they are insensitive to everything else.  We can regard the CC model as
a descriptive model, in the sense that it provides a plausible explanation of what is observed,
but it may prove difficult to implement as a prescriptive model, because in general $T_c$ and $m$ 
are hard to predict theoretically. 

The solid blue curve is the comparable $\tau_{\star}= 5$ Gyr evaporation line for H$_2$O from planets 
scaled from volatile-enriched versions of Earth ($\rho=5.5$, $\Delta M/M=0.01$),
Europa ($\rho=3$ g/cm$^3$, $\Delta M/M=0.1$),
and Ganymede ($\rho=2$, $\Delta M/M=0.5$) --- by chance these three cases are nearly indistinguishable on
this plot, so we show them as one curve.
For these models we set $T_c=T_s$ as, unlike the case for N$_2$-CH$_4$ atmospheres,
 we know of no good reason nor have we seen much evidence to suggest that the upper 
atmospheres of watery worlds should be especially hot or cold.

The blue water line is to the asteroids as the methane line is to the KBOs, whether by accident or design,
but unlike the case for the methane line, which approaches Pluto, Triton, and Titan,
the water line comes nowhere close to explaining the terrestrial planets. 
The water line can be moved into the vicinity of the terrestrial planets by raising $c_{\circ}^2 = k_B T/m$ by an order 
of magnitude. 
This can be done either by converting the H$_2$O to H$_2$ 
or by invoking a hot upper atmosphere.  The result of raising $c_{\circ}^2$ is illustrated
by the dashed blue curve, computed from the same CC model as for water but with $m=2m_{\mathrm{H}}$.  
Such a model may also be relevant to Hayashi-like primary nebular atmospheres,
as it is reasonable to anticipate chemical equilibration and exchange between H$_2$, H$_2$O, and silicates at the surface if the atmosphere
is deep
\citep{Hayashi1979,Sekiya1980a,Sekiya1981,Ikoma2006}.
 
\subsection{The water vapor runaway greenhouse}
\label{section:runaway greenhouse}

The runaway greenhouse threshold is expected to be a weak function of planetary parameters.
To illustrate, assume that the runaway greenhouse limit is set by a troposphere saturated with water vapor becoming optically thick 
\citep{Nakajima1992}.
In the absence of pressure broadening,
optical depth $\tau$ will scale as the column depth;
with pressure broadening this will be multiplied by the pressure to a power $\xi$ on the order of unity \citep{Pollack1969,Robinson2012,Goldblatt2013,Robinson2014}.
For a pure water vapor atmosphere, these considerations imply that \citep{Goldblatt2015}
\begin{equation} \label{tau}
\tau = \kappa \frac{p_{\mathrm{vap}}^{1+\xi} }{ g}, 
 \end{equation}
where $\kappa$ is an opacity.
If we approximate the vapor pressure of water by a simple exponential
\begin{equation}  \label{runaway_scaling_one}
p_{\mathrm{vap}} = p_w \exp{\left( -T_w/T \right)},
 \end{equation}
with $T_w=6000$ K and $p_w=2.1\times 10^7$ bars,
the runaway greenhouse limit should scale as
\begin{equation}
\label{runaway_scaling_two}
I_r \propto T^4 \propto \left( \frac{\left(1+\xi\right) T_w }{ \ln{\left(\kappa p_w^{1+\xi}/g \right)} }\right)^{\!\!4},
 \end{equation}
an expression that is well approximated over the range of interest by
$I_r \propto g^{0.2/(1+\xi) } $.
The surface gravity $g$  
is expressed in terms of $v_{\mathrm{esc}}$ and planet density $\rho$ by
$g^2 \propto \rho v^2_{\mathrm{esc}} $.
This leaves
\begin{equation}
\label{runaway}
 I_r \propto \left(v^2_{\mathrm{esc}}\,\rho \right)^{0.1/(1+\xi) } .
 \end{equation}
This relation is plotted with $\xi=1$ and $\rho=3$ on Figure \ref{Insolation} as
the ``H$_2$O runaway greenhouse.'' 
GCM-based estimates of the onset of the runaway range between 1.1 to 1.7 solar constants,
with details of cloud structure and cloud physics and planetary rotation responsible for much of the variance \citep{Abe2011,Yang2013,Leconte2013a,Leconte2013b,Wolf2015,Way2016}.
The range of uncertainty is encompassed by the thickness of the plotted line.

\subsection{The Moon}

Figure \ref{Insolation} shows the Moon rising above the intersection of the water lines.
In all likelihood the apparent desiccation of the Moon is a 
memory of how the Moon was made rather than the ruin of a more promising world, but 
as plotted here, the Moon as a habitable world appears to be marginally unstable against both thermal escape
and the runaway greenhouse.  
Today, most of the lunar surface is unstable to Jeans escape of water \citep[e.g.,][p.\ 138]{Catling2017}.  
When viewing the Moon from the viewpoint of terraforming it, 
both problems would be made more tractable by providing abundant heavy ballast gases 
to reduce the atmosphere's scale height. 

\subsection{Extrasolar Giant Planets}
\label{section:EGP}

Here we address thermal evaporation of Extrasolar Giant Planets (EGPs).
The topic has been addressed many times \citep[e.g.,][]{Owen2013} with much more sophisticated models than we employ here.
Temperatures in the thermospheres of highly-XUV-irradiated EGPs are expected to be much hotter than the underlying atmospheres \citep[e.g.,][]{Murray-Clay2009,Koskinen2014}.  The high temperature facilitates escape.
We will discuss the XUV-energy-limited escape approximation in Section \ref{section:XUV vs escape velocity} below.
Models of silicate-rich gas giants have also tended to predict warm or even hot upper atmospheres, because
small molecules made of rock-forming elements such as TiO are better absorbers of visible light than they
are emitters of thermal infrared radiation. 
However, early reports of thermal inversions on hot Jupiters \citep[e.g., on HD 209458b;][]{Burrows2007,Knutson
2008} have since been cast into doubt by analyses of more precise data of higher spectral resolution \citep{Line2016}.
The hot thermosphere (and possibly warm stratosphere)
 suggests that escape from the top of the atmosphere is relatively easy,
and therefore that the bottleneck to escape is imposed at some deeper level in the atmosphere.
If, for example, the bottleneck were at the homopause, it would be controlled by turbulent mixing,
which in turn is controlled by the total energy flux moving through the planet (not XUV).   
Here we employ an isothermal approximation that includes the entire atmosphere and the entire
planetary energy budget in escape.

A difference from the icy worlds is that  is that the highly-irradiated EGPs
 are close enough to their stars that tidal truncation must be taken into account.   
This is conveniently done in terms of the Hill sphere distance defined by
\begin{equation}
\label{hill}
r_h = a \left( \frac{M}{ 3 M_{\star}} \right)^{1/3}
\end{equation}
where as above $a$ refers to the star-planet distance and $M_{\star}$ to the mass of the star \citep{Erkaev2007}. 
Along the star-planet axis Eq \ref{force} becomes
\begin{equation}
\label{momentum_2}
u\frac{\partial u }{ \partial r} + \frac{1}{ \rho} \frac{\partial p }{ \partial r} = -\frac{GM }{ r^2} + \frac{GMr }{ r_h^3},
\end{equation}
and Eq \ref{parker} becomes
\begin{equation}
\label{wind_2}
\left(u^2 - c_{\circ}^2\right)\frac{1}{ u} \frac{\partial u }{ \partial r} = \frac{2c_{\circ}^2}{ r} -\frac{GM }{ r^2} + \frac{GMr }{ r_h^3} .
\end{equation}
Equation \ref{wind_2} assumes spherical symmetry for tidal forces, which is not a good assumption.
Its crudeness is probably comparable to treating irradiation as globally uniform or the temperature as isothermal.
The critical point is found by solving the cubic for $r_c$
\begin{equation}
\label{critical_point}
  \frac{2c_{\circ}^2}{ r_c} - \frac{GM }{ r_c^2} + \frac{GMr_c }{ r_h^3} = 0 .
\end{equation}
Equation \ref{wind_2} is easily integrated analytically, 
\begin{equation}
\label{wind_3}
ur^2 = u_c r_c^2 \exp{\left\{ \frac{1}{ 2} \frac{u^2}{ c_{\circ}^2} - \frac{1}{ 2}  \frac{u_c^2}{ c_{\circ}^2} + \frac{GM}{ c_{\circ}^2 r_c} - \frac{GM }{ c_{\circ}^2 r} + \frac{1}{ 2} \frac{GMr_c^2}{ c_{\circ}^2 r_h^3} - \frac{1}{ 2}  \frac{GMr^2}{ c_{\circ}^2 r_h^3}\right\}} .
\end{equation}
Note that $u(r)$ is independent of density.   
Near the surface, where $u^2 \ll c_{\circ}^2$, Eq \ref{wind_3} can be rewritten as an equation for the flow velocity at the surface $u_s$
using the critical point conditions. 
The flux at the surface $\rho_s u_s$ is obtained by multiplying by the surface density $\rho_s$.
\begin{equation}
\label{wind_4}
\rho_s u_s = \rho_s c_{\circ} \frac{r_c^2}{ r_s^2} \exp{\left\{ - \frac{1}{ 2} + \left(1 - \frac{r_s^2}{ r_c^2}\right) \left( \frac{GM}{ 2r_c c_{\circ}^2} -1 \right) + \left(1 - \frac{r_s}{ r_c}\right) \frac{GM}{ r_c c_{\circ}^2} \right\}}
\end{equation}
To use Equation \ref{wind_4} requires choosing $\rho_s$ at the lower boundary using other information.   

In EUV- and XUV-driven escape studies, the lower boundary is typically set at
the homopause or at the base of the thermosphere, with the latter
defined by a monochromatic optical depth in XUV radiation \citep[e.g.,][]{Watson1981,Tian2005,Lammer2013}.
 The homopause can be a reasonable if escape is slow enough that a homopause exists. 
However, as we show below in Section \ref{section:diffusion}, for a gas giant to evaporate in 5 Gyrs, 
the flux of hydrogen to space is too great by orders of magnitude for a homopause to exist.
 (A homopause can still be defined as the altitude where turbulent mixing $K_{zz}$ equals molecular
 diffusion, but it will not manifest as a change in mixing ratios.) 
In a hydrocode simulation of an XUV-heated wind, the lower boundary condition can be justified {\it a postiori}
as self-consistent for a particular model by showing the model to be insensitive 
to changing it \citep[e.g.,][]{Watson1981,Murray-Clay2009,Owen2016}.  

In the simplest picture a gas giant does not have a well-defined surface.
Under these conditions the whole planet takes part in the flow to space, with  
the source of the escaping gas being the shrinking or rarefaction of the interior. 
To first approximation, all EGPs have roughly the same radius;
the typical radius of a hot Jupiter is 84,000 km \citep{Fortney2007}. 
Constant radius is a property of polytropes with a $p = K \rho^2$ equation of state.
The radius is $R = \sqrt{2K/4\pi G}$, where $G$ is the Newtonian gravitational constant \citep{Hubbard1973}. 
Using observed radii of the known roster of transiting planets gives $K=3\pm1\times 10^{12}$ cm$^5$ g$^{-1}$ s$^{-2}$.  
The interior is described by an analytic solution 
 \begin{equation}
 \label{Hubbard}
 \rho(r) = \frac{\pi M }{ 4R^3} \frac {R}{ \pi r} \sin{\left( \frac{\pi r}{ R}\right)} .
  \end{equation}
We set the lower boundary where the inner polytrope and the outer isothermal envelope meet; i.e. where the polytropic pressure equals the pressure of an ideal gas at the planet's effective temperature $T_{\mathrm{eff}}$.  
 This gives 
 \begin{equation}
 \label{LBC}
 \rho_{\rm lbc} = {c_{\circ}^2/K}
 \end{equation}
The other equation pertinent to escape is the global energy balance,
\begin{equation}
\label{total_energy_balance}
 \frac{\pi R^2 L_{\star} }{ 4\pi a^2}= \frac{4\pi R^2 \sigma T^4_s}{ 1-\alpha} + \frac{GM{\dot M}}{ R} \left(1 - \frac{3}{2} \frac{R}{r_h} + \frac{1}{ 2} \frac{R^3}{ r^3_h} \right) 
 + {\dot M} c_p\left( T_c-T_s \right)
\end{equation}
The terms involving ${\dot M}$ are the work done against gravity and any excess heat left in the gas as it escapes.   
As was the case in Section \ref{section:CC} above, if thermal escape is extended over 5 billion years, 
the ${\dot M}$ terms in Eq \ref{total_energy_balance} are negligible,
and Eq \ref{total_energy_balance} reduces to the usual expression for effective temperature,
\begin{equation}
\label{usual effective temperature}
\frac{L_{\star} }{ 4\pi a^2} \approx \frac{4\sigma T^4_{\mathrm{eff}}}{ 1-\alpha} .
\end{equation}
For the EGPs we set the Bond albedo $\alpha=0.1$, $m = 2.4 m_{\mathrm{H}}$, $R_s=8.4\times 10^9$ cm,
and in keeping with the isothermal assumption, we set $T_c=T_s=T_{\mathrm{eff}}$.
The magenta curve in the upper-right-hand region of Figure \ref{Insolation} is computed for complete evaporation of the planet
in $\tau_{\star} =5$ Gyrs;
i.e., we solve Eqs \ref{wind_4}, \ref{LBC}, and \ref {total_energy_balance}
for $v_{\mathrm{esc}}$ such that ${\dot M} \tau_{\star} = M$.
The simple model bounds the population of EGPs (blue disks) rather nicely.
We stress that this is not an XUV-driven escape model.
The planet evaporates because the planet is thermally unstable, not because XUV heating is removing the outer 
atmosphere.

\section{XUV-driven escape}
\label{section:XUV vs escape velocity}

Stellar EUV and X-ray radiation can be very effective at driving the escape of H and H$_2$ from young planets
\citep{Hayashi1979,Sekiya1980a}.
\citet{Urey1952} put it succinctly, ``Hydrogen would absorb light from the sun in 
the far ultra-violet and since it does not radiate in the infra-red [it] would be lost very rapidly.''
\citet{Urey1952} regarded hydrogen escape as obvious and an essential process in planetary evolution (and of habitable Earth in particular),
but he did not quantify it.
\citet{Hayashi1979} proposed that massive hydrogen-rich atmospheres of young planets were removed by 
copious EUV (``extreme ultraviolet,'' $\lambda < 100$ nm) and X-ray ($\lambda < 20$ nm) radiations from young stars \citep{Sekiya1980a,Sekiya1981}. 
Hayashi's idea has proved fruitful and subsequent work on EUV and X-ray driven escape has been
voluminous (see \citet{Tian2015} and \citet{Catling2017} for recent reviews).
EUV and X-ray radiations are usually linked in the literature as XUV radiation because they are expected to be related in stars.
We will use the XUV notation here. 

  \begin{figure}[!htb] 
   \centering
   \includegraphics[width=1.0\textwidth]{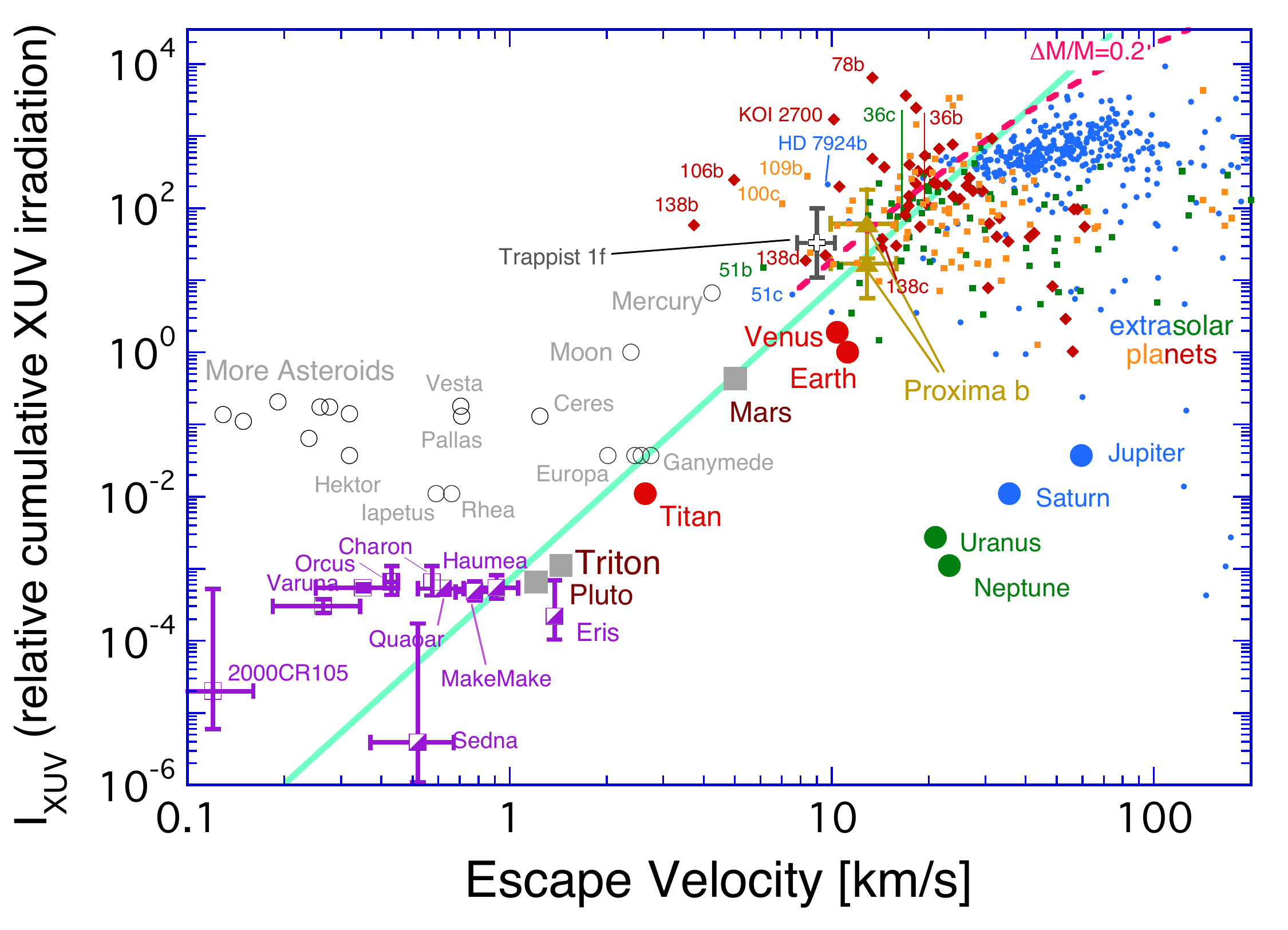} 
   \caption{\small The analog to Figure \ref{Insolation} for estimated cumulative XUV irradiation, which is often
   hypothesized to be the driving force behind planetary evaporation \protect\citep{Tian2015}.
   Uncertainties are much larger here than in Figure \ref{Insolation} because all the XUV fluences need to be reconstructed,
   including the normalizing XUV fluence for Earth.
   No attempt is made to estimate errors for planets outside the solar system.
   The $I_{\mathrm{xuv}} \propto v_{\mathrm{esc}}^4$ line is drawn by eye.
   The dashed magenta curve is for XUV-driven energy-limited escape from tidally-truncated hot EGPs.
   It is labeled by fractional mass lost $\Delta M/M$.
   The model is described in Section \ref{section:XUV-EGP} below.
   Proxima b and Trappist 1f are plotted as described in Sections 6 and 7 below.
 }
\label{XUV}
\end{figure}

In practice it is challenging to test the XUV hypothesis because the bulk of the XUV that a star emits in its lifetime is emitted
when the star is very young, and hence for all but the youngest exoplanets the relevant stellar XUV fluxes are not observables.
Rather, each star's ancient XUV flux needs to be reconstructed from imperfectly known empirical relationships that link stellar age
and spectral type with observed XUV emissions.  
For our purposes, the matter is made fuzzier by the uncertainty that surrounds the fiducial star --- our Sun --- as different extrapolations
differ markedly for the Sun when young.

\citet{Lammer2009} scaled XUV fluxes both with age and with spectral type for F, G, K, and M stars.
They give two part power laws of the general form
$L_{\mathrm{xuv}}  \propto t_{\star}^{-\beta}$, 
with relatively shallow slopes ($\beta < 1$) before 0.6 Gyr and
relatively steep slopes ($\beta >1$) thereafter.
In this prescription the cumulative XUV flux is a well-defined integral dominated by the saturated phase.
We performed these integrals and generalized their result to a simple power law
\begin{equation}
\label{Lammer1}
L_{\mathrm{xuv}} \propto L_{\star}^{0.4}
\end{equation}
which we then express in normalized form for plotting on Figure \ref{XUV} as
\begin{equation}
\label{Lammer2}
I_{\mathrm{xuv}} = \frac{a_{\oplus}^2}{ a^2} \left( \frac{L_{\star}}{ L_{\odot}}\right)^{0.4} = I \left( \frac{L_{\star}}{ L_{\odot}}\right)^{-0.6} .
\end{equation}
The scaling in Figure \ref{XUV} is therefore with respect to a model of the total cumulative XUV radiations
emitted by the ancient Sun, including the early saturated phase.  

Not surprisingly, Figure \ref{XUV} looks a lot like Figure \ref{Insolation}, since only the exoplanets have been changed. 
It is interesting that the EGPs (blue disks) in particular form a tighter distribution, 
which may be a hint that with XUV we are on the right track.
The $I_{\mathrm{xuv}} \propto v_{\mathrm{esc}}^4$ line is drawn in by hand to guide the eye.
The dashed magenta curve represents the quantitative predictions of a basic XUV-driven escape
model to be described below in Section \ref{section:XUV-EGP}. 
The XUV-driven escape model works rather well for EGPs.

\section{XUV-driven escape: Part II}
\label{section:three}

In this section we address the expected form an energy-limited power law would take,
and compare its predictions to those of XUV driven escape (Figure \ref{XUV_vs_x}).
  
 \subsection{General considerations}
\begin{figure}[!htb] 
   \centering
   \includegraphics[width=1.0\textwidth]{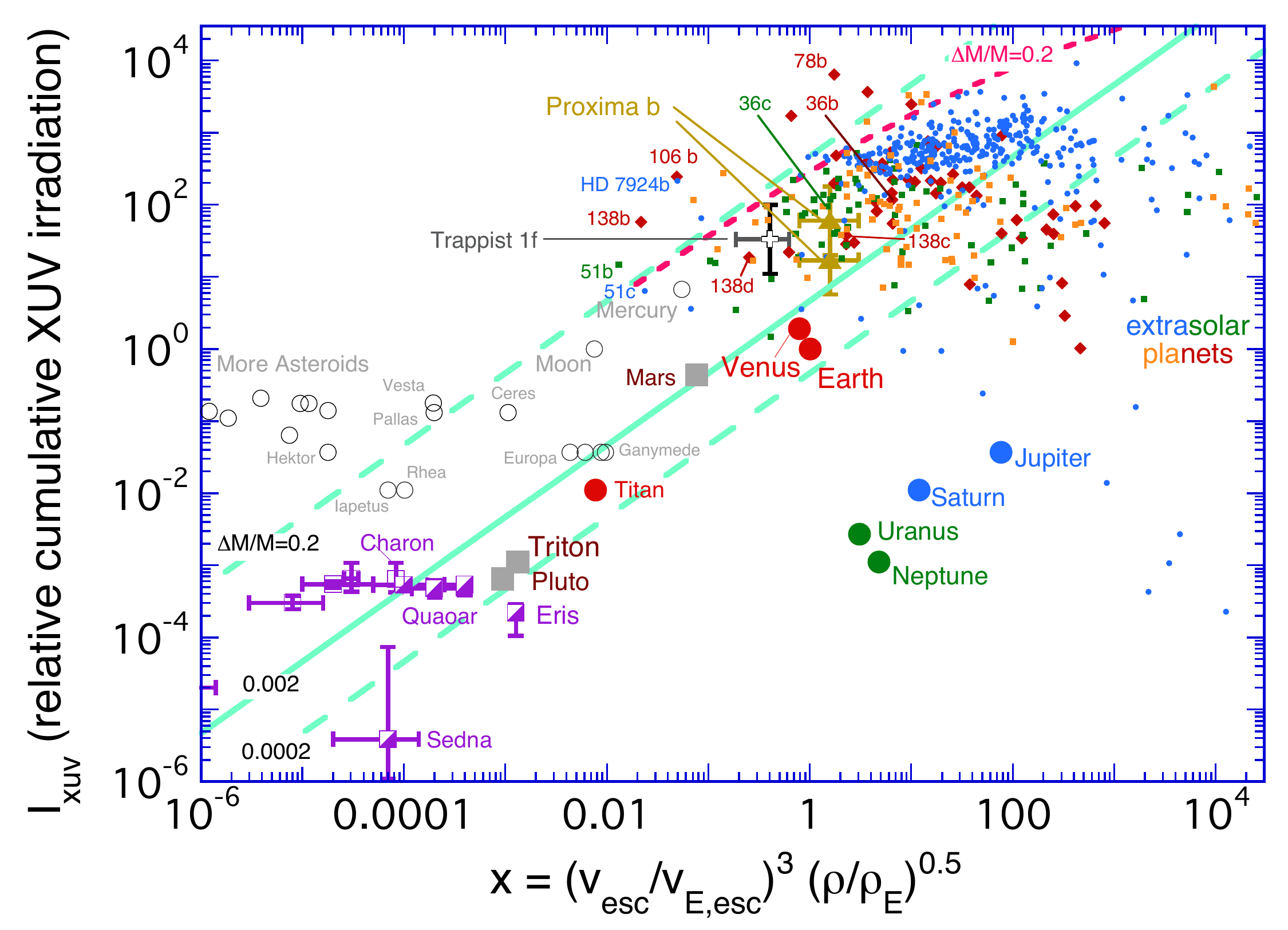} 
   \caption{\small Here the data from Figure \ref{XUV} are plotted against the expectations of the simplest theory. 
   Energy-limited models predict that, if tidal stripping is unimportant, 
   the insolation $I_{\mathrm{xuv}}$ should be linearly proportional to the quantity $x \equiv v_{\mathrm{esc}}^3 \sqrt{\rho} $.
   Here we express $x$ as normalized to Earth, $\left(v_{\mathrm{esc}}/v_{\oplus\mathrm{esc}}\right)^3 \sqrt{\rho/\rho_{\oplus}}$,
   so that Earth sits at $(1,1)$.
   The green diagonal lines represent a family of these predictions, denoted on
  the plot by the relative fraction $\Delta M_{\mathrm{el}}/M$ of the planet's mass that can be lost in XUV-driven escape.
   The upper green line $\Delta M/M=0.2$ is the extension of the dashed magenta curve for XUV-driven energy-limited escape from tidally-truncated hot EGPs.  
  The middle (solid) green line $\Delta M/M = 0.002$ approximates the upper bound ($f(\mathrm{H}_2)=1$) on diffusion-limited
  escape of H$_2$ for systems that are a few billion years old.  
  The lower (dashed) green line $\Delta M/M = 0.0002$ approximates the upper bound on diffusion-limited
  escape of H$_2$ if rapid escape is restricted to young, XUV active stars. 
 Proxima b and Trappist 1f are plotted as described in Sections 6 and 7 below.
}
\label{XUV_vs_x}
\end{figure}

It is possible to quantify the predictions of the XUV hypothesis if the escape is energy limited.
Energy-limited escape is expected if the XUV radiation is too great 
for the incident radiation can be thermally conducted to the lower atmosphere \citep{Watson1981}).
If tidal truncation is for the moment neglected, the energy-limited escape can be expressed as 
\begin{equation}
\label{energy limit no Hill}
{\dot M_{\mathrm{el}}} = \frac{\eta \pi R^3 L_{\mathrm{xuv}}(t) }{ 4\pi a^2 GM }
\end{equation}
where ${\eta}$ is an efficiency factor that is usually taken to be $0.1 < {\eta} < 0.6$ \citep[e.g.,][]{Lammer2013,Owen2013,Koskinen2014,Bolmont2017}. 
The mass loss efficiency $\eta$ is less than the heating efficiency (fraction of incident XUV energy converted to heat) because the escaping gas is
hotter, more dissociated, and more ionized than it was before it was irradiated. 
The factor $\eta$ is a function of $T$, being smaller in cooler gas ($< 3000$ K) in which H$_3^+$ is a major radiative coolant \citep{Koskinen2014} and smaller in hot gas ($\sim 10^4$ K) in which collisionally excited Lyman $\alpha$ is a major coolant \citep{Murray-Clay2009}. 
 In a steam atmosphere, FUV (``far ultraviolet,'' $100 < \lambda < 200$ nm) can also be important because it is absorbed by H$_2$O, O$_2$, and CO$_2$, provided irradiation is modest enough to leave the molecules intact \citep{Sekiya1981}.  The contribution of FUV is implicitly folded into $\eta$.
The total mass loss 
\begin{equation}
\label{Delta M}
\Delta M_{\mathrm{el}} = \int^{\tau_{\star}}_0 {\dot M_{\mathrm{el}}}\, dt 
\end{equation}
 is obtained by integrating the star's XUV radiation history expressed in terms of 
the Sun's history using 
$I_{\mathrm{xuv}} = \left({L_{\mathrm{xuv}}/L_{\mathrm{xuv}\odot}}\right)\left(a_{\oplus}/ a\right)^2$.
For the XUV history of the Sun itself we follow \citet{Ribas2005,Ribas2016}.
\begin{eqnarray}
 \label{XUV_history}
  L_{\mathrm{xuv}\odot} & = L_x \left( t / t_x \right)^{-\beta} & \qquad t > t_x \nonumber \\
  L_{\mathrm{xuv}\odot} & = L_x  \phantom{\left( t / t_x \right)^{-\beta}} & \qquad t < t_x 
 \end{eqnarray}
where $t_x=0.1$ Gyr, $\beta=1.24$, and where
\begin{equation}
\label{saturated}
L_x = 1.6\times 10^{30} \left( t_x / t_{\odot} \right)^{-\beta} \quad \mathrm{ergs~}\mathrm{s}^{-1}
\end{equation}
is the saturated upper bound on the Sun's youthful excess.
Cumulative escape is then 
\begin{equation}
\label{cumulative escape}
\frac{\Delta M_{\mathrm{el}} }{ M} = \frac{\eta L_x t_x }{ a^2_{\oplus} }\sqrt{\frac{3G}{ 8\pi}} \frac{I_{\mathrm{xuv}}}{ v_{\mathrm{esc}}^3\sqrt{\rho}} \frac{1 }{ \beta -1}\left( \beta - \left(t_x/\tau_{\star}\right)^{\beta - 1}\right)
\end{equation}
which can be rearranged as a linear relation between $I_{\mathrm{xuv}}$ and $x$
\begin{equation} \label{Ixuv vs x}
I_{\mathrm{xuv}} = x \sqrt{\frac{8\pi }{3G}} \frac{\Delta M_{\mathrm{el}} }{ M} \frac{a^2_{\oplus} \left(\beta -1\right) }{ \eta L_x t_x \left( \beta - \left(t_x/\tau_{\star}\right)^{\beta - 1}\right) } 
\end{equation}
with $x$ defined by
\begin{equation} \label{x definition}
x \equiv {v_{\mathrm{esc}}^3}\sqrt{\rho} .
\end{equation}
Results are plotted in terms of the parameter $x$ on Figure \ref{XUV_vs_x} for $\eta=0.2$ as green lines 
and labeled for a range of lost masses $2\times 10^{-4} \leq {\Delta M_{\mathrm{el}} /M} \leq 0.2$.  
As in Figure \ref{XUV}, we apply $L_{\mathrm{xuv}} \propto L_{\star}^{0.4}$ to the data in order to present a single relation
that spans all the planets.  
The Solar System is fit by ${\Delta M_{\mathrm{el}} /M} \approx 0.001$,
a trend that does not extend to the exoplanets, which are better matched by ${\Delta M_{\mathrm{el}} /M} \approx 0.1$.

\subsection{Diffusion-limited escape}
\label{section:diffusion}

The diffusion-limited flux gives the upper bound on how quickly hydrogen can selectively escape
by diffusing through a heavier gas that does not escape; more properly, it is the upper
limit on the difference between hydrogen escape and heavy-constituent escape.
It is often thought of in the context of vigorous hydrogen escape driven by XUV radiation \citep{Sekiya1980b,Zahnle1986,Hunten1987},
but it is quite general \citep{Hunten1976}.
The diffusion-limited flux regulates hydrogen escape or 
H$_2$ abundance on Venus, Earth, Mars, and Titan today \citep[see review by][Chapter 5]{Catling2017}. 
It is likely that, should the diffusion-limited flux be smaller than the energy limit,
the H$_2$ mixing ratio $f(\mathrm{H}_2)$ will increase until the two limits are equal, as seen on Titan and Mars.
But if $f(\mathrm{H}_2) \rightarrow 1$ and the energy limit still exceeds the diffusion limit, it is not obvious what happens.
The atmosphere may either escape as a whole (although the heavier gases escape more slowly, so that the remnant atmosphere
becomes mass fractionated), or hydrogen escape is throttled to the diffusion limit and the excess energy is radiated to space
by the heavy gases. 
A possible example of escape at the diffusion limit among the EGPs is HD 209458b \citep{Vidal2003,Vidal2004,Yelle2004,Koskinen2013}.

The upper bound on the escape flux of a gas species of molecular mass $m_i$ from a static gas atmosphere of molecular mass $m_j$ ($m_i < m_j$) in the diffusion limit from an isothermal atmosphere is 
\begin{equation}
\label{diffusion limit}
\phi_{i,\mathrm{dl}} = f_i \frac{GM\left(m_j-m_i\right) }{ R^2} \frac{b_{ij}}{ k T},
\end{equation}
where $f_i$ is the mixing ratio of the light gas
and $b_{ij}$ is the binary diffusion coefficient between the two species $i$ and $j$.
Typically $b_{ij} \propto T^{0.75}$.
The corresponding mass loss rate is 
\begin{equation}
\label{diffusion limit mass loss}
{\dot M} = 4\pi R^2 m_i\phi_{i,\mathrm{dl}} .
\end{equation}
The timescale for losing an atmosphere of mass $\Delta M_{\mathrm{dl}}$ is $\tau_{\mathrm{dl}} = \Delta M_{\mathrm{dl}}/{\dot M}$.
The upper bound that diffusion puts on atmospheric escape can be written  
\begin{equation}
\label{diffusion upper bound}
\frac{\Delta M_{\mathrm{dl}} }{ M} = \frac{4\pi m_i f_i\left(m_j-m_i\right) b_{ij} G \tau_{\mathrm{dl}} }{  kT} .
\end{equation}
It is notable that this ratio is very nearly independent of planetary parameters - i.e., this constraint is the same for all planets.
For H$_2$ escaping through N$_2$, $b_{ij} \approx 1.5\times 10^{19} \left(T/300\right)^{0.75}$,
for which
\begin{equation}
\label{diffusion Delta M}
\frac{\Delta M_{\mathrm{dl}} }{ M} = 0.001 f(\mathrm{H}_2)  \left({1000/T}\right)^{0.25} \tau_{\mathrm{Gyr}}   
\end{equation}
with $\tau_{\mathrm{Gyr}}$ measured in Gyrs.  
This means that it is difficult for a planet to selectively lose more than about 0.5\% of its mass as H$_2$ in 5 Gyr,
even if H$_2$ is a major constituent ($f(\mathrm{H}_2) \rightarrow 1$), as it often appears to be on exoplanets. 
In many XUV-limited escape scenarios the time available for energy-limited escape is less than a few hundred million years,
which reduces the maximum differential H$_2$ loss to less than 0.05\% of the planet's mass. 
Whether this is an important constraint depends on several factors.
If H$_2$ is overwhelmingly abundant and the heavy gases are inefficient radiative coolants,
they can be carried along, and $\phi_{i,\mathrm{dl}}$ becomes the difference between H$_2$ escape and heavy gas escape \citep{Sekiya1981,Zahnle1986}.
 If the heavy gases condense, they can be separated from H$_2$ by precipitation and
 the gas diffusion limit does not apply.
But for warm planets with considerable reservoirs of volatiles other than H$_2$, the constraint may set
the boundary between planets that evolve to a vaguely Earth-like state vs.\ those that never 
progress past a vaguely Neptune-like state.
That XUV-driven escape should lead to such a bimodal distribution of planets has also been made 
on the basis of the limited XUV energy available \citep{Owen2013}.   

\subsection{Extrasolar Giant Planets}
\label{section:XUV-EGP}

The simple linear relation between $I_{\mathrm{xuv}}$ and $x$ in Equation \ref{Ixuv vs x}
does not apply for the close-in planets that are afflicted by tidal truncation.  
For these we need to include the Hill sphere terms, which break the $I_{\mathrm{xuv}} \propto x$ relation.
For these planets we start with a tidally truncated XUV-heated energy-limited escape flux,
\begin{equation}
\label{XUV-heated energy-limited escape flux}
\frac{\pi R^2 \eta L_{\mathrm{xuv}} }{ 4\pi a^2} = \frac{GM{\dot M_{\mathrm{el}}} }{ R}\left(1 - \frac{3}{ 2} \frac{R}{r_h} + \frac{1}{ 2} \frac{R^3}{ r_h^3}\right) 
\end{equation}
As above, we treat EGPs as all having the same radius $R$.
With $R$ held constant, $M$ can be replaced by $v_{\mathrm{esc}}$; the star-planet distance $a$ can be replaced by $I_{\mathrm{xuv}}$;
and $M_{\star}$ is replaced $M_{\odot}$ by
expressing $L_x/L_{x\odot} \propto \left(L_{\star}/L_{\odot}\right)^{0.4} \propto \left(M_{\star}/M_{\odot}\right)^{1.5}$,
in which the stellar mass-luminosity relationship is conveniently written in the form $L_{\star}\propto M_{\star}^{3.75}$.
With these relations, the $L_x/L_{x\odot}$ ratio cancels out of Eq \ref{XUV-heated energy-limited escape flux}.
The resulting expression between $I_{\mathrm{xuv}}$ and $v_{\mathrm{esc}}$ should hold approximately
for all main sequence stars and their giant planets,
\begin{eqnarray}
\label{I_xuv vs vesc with Hill}
\frac{a_3 I_{\mathrm{xuv}} }{ v_{\mathrm{esc}}^4} & 
= & \frac{\Delta M_{\mathrm{el}} }{ M} \left(1 - a_1 \frac{I_{\mathrm{xuv}}^{1/2} }{ v_{\mathrm{esc}}^{2/3}} 
     + a_2 \frac{I_{\mathrm{xuv}}^{3/2} }{P v_{\mathrm{esc}}^{2}} \right)\\
     a_1 & = & \frac{3}{ 2} \frac{R}{P a_{\oplus}}\left( \frac{6GM_{\odot}}{ R}\right)^{1/3} \nonumber \\
     a_2 & = & \frac{1}{ 2} \frac{R^3}{ a_{\oplus}^3} \frac{6GM_{\odot}}{ R} \nonumber \\
     a_3 & = & \frac{GR}{ a_{\oplus}^2} \frac{\eta L_xt_x }{ \left(\beta -1\right)}\left( \beta - \left(t_x/\tau_{\star}\right)^{\beta - 1} \right) .\nonumber
\end{eqnarray}
Equation \ref{I_xuv vs vesc with Hill} is readily solved for $I_{\mathrm{xuv}}$ as a function of $v_{\mathrm{esc}}$.
Results are plotted for ${\Delta M_{\mathrm{el}} /M} = 0.2$ with $\eta=0.2$
as the dashed magenta lines on Figures \ref{XUV} and \ref{XUV_vs_x}.
For the particular case with $R$ held constant, $x \propto v_{\mathrm{esc}}^4$, so the curves are uniquely defined on both plots.
As has been pointed out by others \citep[e.g.,][]{Owen2013}, 
the quantitative predictions made by the simple XUV model are good enough to be intriguing.
The energy-limited escape and the isothermal escape model 
make similar predictions (compare Figures \ref{Insolation} and \ref{XUV}), but both models depend on tidal forces to truncate the atmosphere, which suggests that tidal truncation is the true control.
 If so, a more accurate description of the tidal potential would be a useful direction to take further research.

\section{Impact erosion}
\label{section:four}

  \begin{figure}[!htb] 
   \centering
   \includegraphics[width=1.0\textwidth]{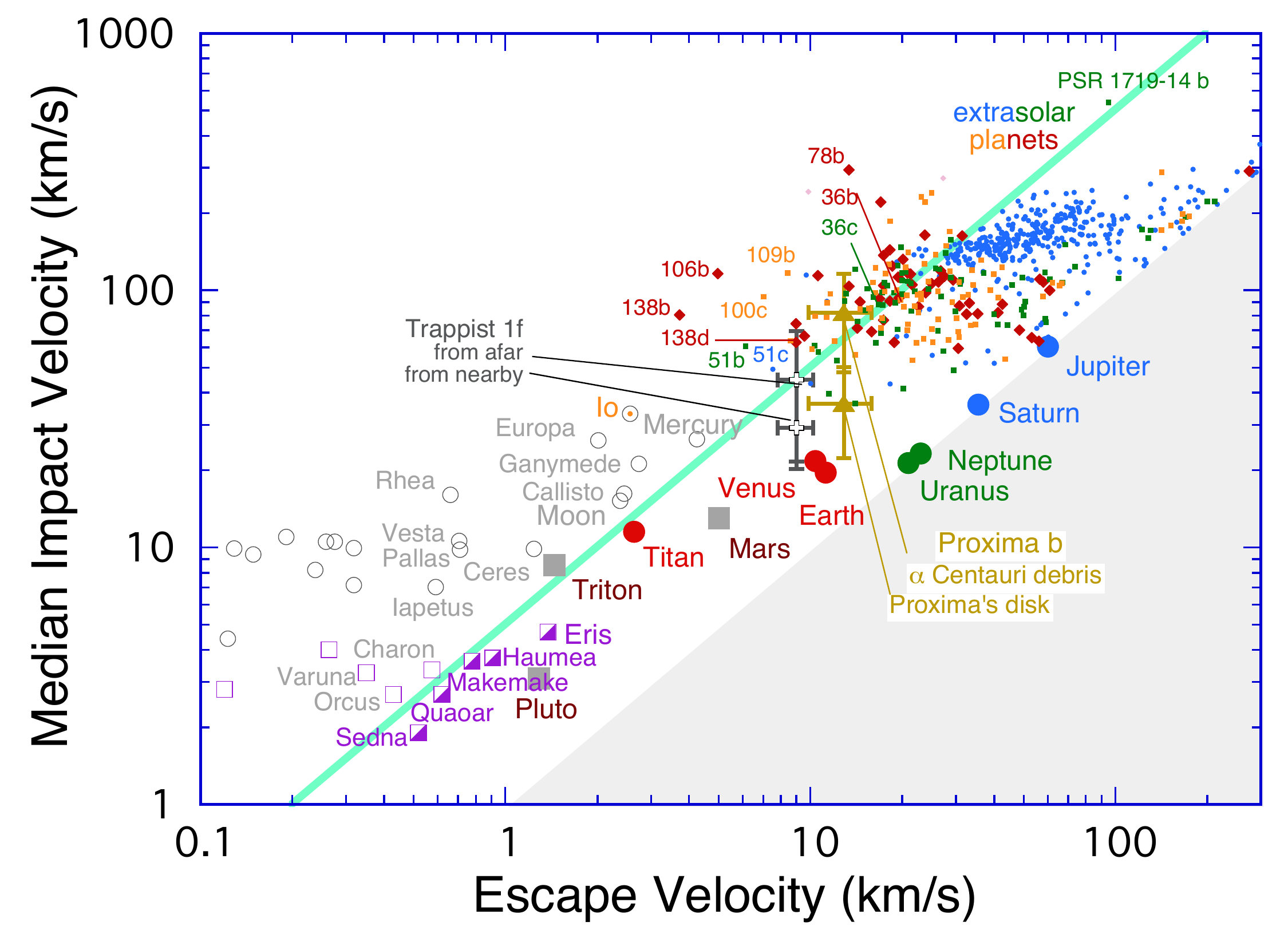} 
   \caption{\small Here typical impact velocities, estimated from the orbital velocities $v_{\mathrm{orb}}$ of the planets,
    are plotted against $v_{\mathrm{esc}}$.
   The shaded area in the lower right is unphysical.
   Solar System bodies with atmospheres, such as Earth, are plotted in solid colors.
   Bodies in the Solar System that are devoid of atmospheres are plotted with open gray symbols.
   Kuiper Belt Objects are purple.
   Transiting exoplanets are plotted in blue disks (Saturns and Jupiters), green boxes (Neptunes), and red diamonds (Venuses).
   Expansive error bars are omitted for clarity.
   The empirical impact erosion stability limit for solar system atmospheres is roughly $v_{\mathrm{imp}}/v_{\mathrm{esc}} = 5$.
   Proxima b (in gold) and Trappist 1f (charcoal) are plotted twice. 
    The higher impact velocity presumes bodies from distant orbits; with Proxima this could mean bodies that
    orbit $\alpha$ Centauri.
   The lower impact velocity presumes bodies in prograde orbits roughly coplanar with the planets.
   Low impact velocities might be expected in the Trappist 1 system.
 }
\label{Impacts}
\end{figure}

Impact erosion of planetary atmospheres can be another path to ruin 
\citep{Walker1986,Melosh1989,Zahnle1992,Zahnle1993,Zahnle1998a,Griffith1995,Chen1997,Brain1998,Newman1999,
Genda2003,Genda2005,Catling2009,deNiem2012,Catling2013,Schlichting2015}.
The basic idea is that a portion of a planetary atmosphere is blasted into space if an impact
is big enough and energetic enough.
Once in space the noncondensing volatiles are presumed dispersed by radiation pressure or the solar wind, whilst condensing materials
are for the most part swept up again.
In detail, how impact erosion actually works remains a work in progress. 
It may be that impact erosion is mostly caused by very large
collisions that drive off much of the atmosphere in a single blow \citep{Korycansky1992,Chen1997,Genda2003,Genda2005},
 or it may be more like sandblasting, with tens of thousands of small impacts each doing a little 
\citep{Walker1986,Melosh1989,Zahnle1992}, or it could be something in between, or a combination of all these effects.
Moreover, impact erosion cannot be evaluated while not also evaluating the impact delivery of new volatiles. 

In previous work, we suggested that impact erosion is likely to be dominated by numerous relatively small projectiles striking the planet's surface
at velocities well in excess of the escape velocity, whilst impact delivery of new volatiles
is likely to be dominated by a few slow-moving, very large volatile-rich bodies \citep{Zahnle1992,Griffith1995}.
Consequently, although the loss of atmosphere by impacts may be plausibly approximated by a continuous function, impact delivery
of volatiles is likely to be profoundly stochastic.
If this is how it works, impact erosion, when it gets the upper hand,
 will annihilate the atmosphere, because as the atmosphere
thins the eroding projectiles become ever smaller and more numerous.
This kind of impact erosion is a good candidate for creating a nearly airless world like Mars \citep{Melosh1989,Zahnle1993}
and it readily accounts for the sharp distinction between the atmospherically-gifted Titan on one hand and the airless Callisto and Ganymede
on the other \citep{Zahnle1992,Griffith1995,Zahnle1998a}. 
But where by chance a single late great impact delivers an atmosphere so massive 
that all subsequent impacts are insufficient to remove it, a considerable atmosphere can be 
left on a planet where one might not expect to find one \citep{Griffith1995}.  
If impact erosion is the chisel that sculpts extra-solar systems, 
we would expect that by chance there will exist a few small, close-in planets
enveloped in appreciable atmospheres.  This may be germane to assessing Proxima b.

However it happens, it is plausible that the efficiency of impact erosion should scale as $v_{\mathrm{imp}} \propto v_{\mathrm{esc}} $,
and therefore in Figure \ref{Impacts} we plot the planets on the grid $v_{\mathrm{imp}}$ {\it vs.} $v_{\mathrm{esc}} $.
To prepare Figure \ref{Impacts} we use solar system impact velocities from \citep{Zahnle2003}
with appropriate updates for the KBOs.
For the exoplanets we assume that the impacting bodies
come from prograde orbits of modest inclination and eccentricity that 
generically resemble those of the asteroids and Jupiter-family comets that strike Earth and Venus.
In the inner Solar system, encounter velocities 
$v_{\mathrm{enc}}$ are typically on the order of $0.5-1.0\times$ the orbital velocity,
with the higher encounter velocity appropriate to matter falling from greater heights above the Sun (i.e., comets) \citep{Bottke1995}.
The circular orbital velocity $v_{\mathrm{orb}}$ of an extrasolar planet 
is computed using the reported period and either the semimajor axis, such that 
$v_{\mathrm{orb}} = 2\pi a/P$,
or the star's mass $M_{\star}$, such that $v_{\mathrm{orb}}^3 = 2\pi GM_{\star}/P$.
 For most planets $a$ and $M_{\star}$ are both listed; for these we take the average. 
What is actually plotted on Figure \ref{Impacts} uses 
\begin{equation}
v_{\mathrm{imp}}^2 = v_{\mathrm{enc}}^2 + v_{\mathrm{esc}}^2
\end{equation} 
with $v_{\mathrm{enc}} = v_{\mathrm{orb}}$.

Although the data are very uncertain, Figure \ref{Impacts} shows clearly that impact erosion cannot be lightly dismissed.
But we are not currently in a position to make quantitative predictions comparable to those we made for insolation-driven escape:
we don't know the actual impact velocities anywhere other than in our own solar system, and even here we meet
with considerable dispersion; nor do we know the volatile contents of the impacting bodies; nor do we have a robust
theory of how impact erosion works; nor have we a robust theory to describe the retention and loss of the delivered volatiles. 
What we can say is that impact erosion has promise as a global explanation,
and because its effects are roughly parallel to those of insolation-driven escape, the two processes might often work together.
For what it is worth, the empirical dividing line for the ensemble, $v_{\mathrm{imp}} \approx 4\!-\!5\, v_{\mathrm{esc}}$, is at a higher $v_{\mathrm{imp}}$ than the $v_{\mathrm{imp}} \approx 2.5\, v_{\mathrm{esc}}$ that had been discussed for Mars \citep{Melosh1989,Zahnle1993}.  
It may be germane that the ejecta from comet Shoemaker-Levy 9 were launched at 20-25\% of the impact velocity \citep{Zahnle1996},
which suggests to the optimist that the relation $v_{\mathrm{imp}} \approx 4\!-\!5\, v_{\mathrm{esc}}$ may hold generally for airbursts in deep atmospheres. 
This in particular is a topic that should be addressed in future work.

\section{Proxima b: On the Beach}
\label{section:five}

In 2016, a planet somewhat more massive than Earth was discovered orbiting the Sun's nearest neighbor every 11.2 days \citep{Anglada2016}.
The planet lies within the conventional habitable zone, in that it intercepts a total amount of insolation comparable to what
the Earth intercepted during its inhabited Archean Eon ca 3 Ga.
 This mix of qualities --- the nearest exoplanet, vaguely Earth-massed, in the habitable zone --- almost guarantees that 
 Proxima b will be explored by humans or their descendants at some point in the distant future.
There has been a fair amount written about Proxima b that does not all need to be repeated here
\citep{Anglada2016,Davenport2016,Ribas2016,Turbet2016,Barnes2017,Coleman2017,Goldblatt2017,Meadows2017}. 
Here we wish to document how we plotted Proxima b on Figures 1-4 and then, briefly, speculate about its habitability.

\subsection{Proxima b: escape velocity}

Proxima b's escape velocity is uncertain because we do not know its radius;
 we do not know whether it is a globe of air, water, earth, or metal. 
 Radial velocity gives $M\sin{i}=1.27 M_{\oplus}$.
We presume the median nominal mass of $M=1.27/\sin{60^{\circ}}=1.47 M_{\oplus}$.
If rocky, using $M \propto R^{3.7}$ \citep{Zeng2016}, we estimate $v_{\mathrm{esc}}=12.7$ km s$^{-1}$. 
To set a rough upper uncertainty we take a more face-on $\sin{30^{\circ}}$ orbit,
for which $v_{\mathrm{esc}}=15.3$ km s$^{-1}$.  To set a rough lower uncertainty, 
we presume that Proxima b is ice-rich with a bulk density half that of a rocky world of the same mass, for which $v_{\mathrm{esc}}=10.2$ km s$^{-1}$.

\subsection{Proxima b: XUV heating}

For Proxima specifically, \citet{Ribas2016} estimate both the current and the cumulative relative XUV irradiations of Proxima b and Earth.
For the present, they estimate that ${I_{\mathrm{xuv}}/I_{\oplus}} \approx 60$.
For the cumulative total, they estimate that ${I_{\mathrm{xuv}}/I_{\oplus}} \approx 16$.
We plot both estimates on Figures \ref{XUV} and \ref{XUV_vs_x}, each with a factor 3 uncertainty.
On Figure \ref{XUV}, Proxima b appears relatively vulnerable to XUV-driven escape.
Only a big and dense Proxima b orbiting an XUV-quiet Proxima plots with the terrestrial
planets of our solar system.  Otherwise, Proxima b plots with the least of the extrasolar Neptunes and a few other smaller planets.
But when the ensemble is replotted according to the expectations of energy-limited flux, 
Proxima b looks rather ordinary (Figure \ref{XUV_vs_x}).
Figure \ref{XUV_vs_x} also suggests that Proxima b at 0.05 AU has intercepted enough XUV energy to drive off
about 1\% of its mass; a waterworld Proxima b should be durable to XUV radiation
if it could be made in the first place, but a more Earth-like hydrosphere could be vulnerable to being wholly lost.

\subsection{Proxima b: insolation}

After it formed, Proxima is presumed to have slowly faded to the main sequence like any small M dwarf.
Figure \ref{Insolation} shows insolation levels from \citet{Ribas2016} at 3 times: when Proxima was just 10 Myr old,
when Proxima was 100 Myr old, and today.  
With respect to total insolation, on Figure \ref{Insolation} Proxima b is much like Earth or Venus.
As \citet{Ribas2016} and \citet{Barnes2017} and many others have pointed out,
when Proxima was young, insolation exceeded the runaway greenhouse threshold at 
Proxima b's current 0.05 AU distance for the first $\sim 150$ Myrs or so of their mutual existence \citep{Barnes2017}.
This means that Proxima b, if it had water when young, would have held it initially in the form of steam,
which eliminates the cold trap as a bottleneck to hydrogen escape.
(Another way to eliminate the cold trap as a bottleneck to escape is to make water vapor a major constituent.
This can happen if other gases are scarce \citep{Wordsworth2013}, or the planet could enter a ``moist greenhouse'' state \citep{Kasting1988,Kasting2015,Leconte2013a,Wolf2015}.)
If hydrogen escape is restricted to the runaway greenhouse epoch, 
estimates of the total XUV-driven energy-limited escape
range from less than an Earth ocean of water \citet{Ribas2016} to 3-10 oceans \citep{Barnes2017},
a difference that can be attributed to assumptions about Proxima as a young star.

\subsection{Proxima b: oxygen}

It has been suggested that, if the source of escaping hydrogen is water, O$_2$ might build up
in the atmosphere at the diffusion-limited rate, and if the hydrogen from several oceans of water escaped, it might be possible for
hundreds of bars of O$_2$ to accumulate in the atmosphere left behind \citep{Luger2015,Schaefer2016,Barnes2017}. 
However, these models do not account for atmospheric photochemical reactions between oxygen and hydrogen that can reduce
the H$_2$ mixing ratio so that $f(\mathrm{H}_2) \ll 1$ and thus throttle hydrogen escape.
For example, the hydrogen escape rate from Mars is currently very slow because the strong negative feedback between oxygen
and hydrogen ensures that both are lost from the atmosphere in the 1:2 ratio of the parent molecule \citep{Hunten1976}.
The same 1:2 ratio holds for oxygen and hydrogen escape from Venus today \citep{Fedorov2011}. 
Moreover, iron in a vigorously convecting mantle has the capacity to consume thousands of bars of O$_2$. 
For example, \citet{Gillmann2009} and \citet{Hamano2013} dispose of the excess oxygen generated by hydrogen escape from Venus's accretional steam atmosphere by putting it into the mantle while still mostly molten under a steam atmosphere.
In any event O$_2$ on Venus has yet to be detected \citep{Fegley2014}.
\citet{Schaefer2016} include aspects of the kinetics of the mantle sink.
If the XUV is very large, hydrogen escape can in principle be vigorous enough to drag the oxygen liberated by water photolysis 
into space \citep{Zahnle1986,Luger2015,Schaefer2016}. 
For example, \citet[][Fig.\ 8]{Zahnle1986} showed that for conditions germane to a steam atmosphere on Venus, molecular diffusion ensures
that oxygen escape must exceed the surface sink on oxygen, no matter how efficient the latter.
\citet{Schaefer2016} reached a similar conclusion for GJ 1132b and other worlds.
When hydrogen escape exceeds the diffusion limit, the diffusion limit becomes the rate
that oxygen is left behind to oxidize the planet or accumulate in the atmosphere.    
However, in addition to not including atmospheric chemistry,
  neither model took into account that because the mixed wind is heavier, it must be hotter than pure hydrogen and thus
has more power to cool itself radiatively.
   
\subsection{Proxima b: impacts and impact erosion}

The history and nature of impacts experienced by Proxima b are almost wholly conjectural \citep{Coleman2017}.
Still, impacts happen.
It may be helpful to divide impactors into three general classes: 
(i) material co-orbital with Proxima b;
(ii) material in orbits about Proxima (analogous to the Sun's asteroid and Kuiper belts)
 and (iii) material in orbits about $\alpha$ Centauri A or B or both (Kuiper belts and Oort clouds would be Solar System analogs).
 The first category is swept up very quickly and is better regarded as part of Proxima b's accretion and will be addressed separately below.
 In the second category we imagine bodies in prograde orbits roughly coplanar with Proxima b and perturbed from relatively
 distant orbits, with aphelia at say 1 AU, into highly elliptical Proxima b-crossing orbits.
 By analogy to impacts on Earth \citep{Bottke1995},
 we estimate that typical encounter velocities would be of the order of $\langle v_{\mathrm{enc}} \rangle \approx 0.5-0.8 \,v_{\mathrm{orb}} $,
 which corresponds to $v_{\mathrm{imp}} \approx 30 \pm 15$ km s$^{-1}$.
 The third category is analogous to the comets and asteroids that strike the Galilean satellites (in mildly hyperbolic orbits with respect to Juoiter),
 with almost all of the velocity of the stray body attributable to the gravitational well of the central body. 
 We have previously modeled this scenario for the Galilean satellites,
 taking into account the distribution of impact probabilities associated with the distribution of encounter orbits \citep{Zahnle1998b}.
 Generalizing from \citet{Zahnle1998b}, we estimate that close encounters of the third kind would fall in the range
 $v_{\mathrm{enc}} \approx 1.4-2.4 \,v_{\mathrm{orb}}$; 
  i.e., we estimate that $v_{\mathrm{enc}} \approx 90 \pm 25$ km s$^{-1}$.
Cases (ii) and (iii) are plotted on Figure \ref{Impacts}.
It is apparent at a glance that Proxima b is more vulnerable to the negative consequences of impacts than are
Earth and Venus, a not surprising observation that has been anticipated \citep[cf,][]{Raymond2007,Lissauer2007}. 
We conclude that almost all the collisions that matter to impact erosion and impact delivery must be from 
debris orbiting Proxima itself, at velocities that are marginally more erosive than what we see in the inner Solar System.  

\subsection{Proxima b: accretional heating}

If averaged over 100 Myrs, the energy of accretion of a planet like Earth is comparable in magnitude to the insolation over that same period.
This was very important for Earth and Venus because they likely accreted on a 30-100 Myr timespan,
and the added energy of accretion pushed both planets above their runaway greenhouse limits \citep{Matsui1986,Abe1988,Zahnle1988,Hamano2013}. 
For Earth, the steam atmospheres were episodic transients after big impacts, but Venus's steam atmosphere was probably irreversible,
and hence led directly to the profound desiccation of Venus's atmosphere and mantle \citep{Hamano2013}.

By contrast to Earth and Venus, Proxima b could have accreted very quickly.
 \citet{Lissauer2007} showed that in basic Safronov accretion theory,
 HZ (habitable zone) planets of small M dwarfs are expected to accrete in less than $10^5$ years,
orders of magnitude faster than Earth or Venus.
Accretion in $10^5$ years implies a surface temperature of 2000 K if airless and perhaps 4000 K if the planet had an
atmosphere, which at temperatures like these it most certainly would have had \citep{Lupu2014}.
In its potential for rapid accretion Proxima b is more like a Galilean satellite than a Solar System planet.
 The comparable accretion time for Europa is 200 years, which scarcely seems credible
 for a small world that retains a lot of water, to say nothing of icy Callisto accreting in just 3000 years yet never fully melting.
 Evidently Europa's and Callisto's accretions were governed by the supply of new matter from the Sun's accretion disk to Jupiter's accretion disk,
 rather than by the properties of Jupiter's accretion disk.
 The ruling time scale then becomes that of forming the solar system as a whole, 
 which appears to have been on the order of 3 million years (and slow enough to preserve a cold Callisto).
But Proxima b is much bigger than Europa or Callisto.
Even if material were supplied to the Proxima system from an unknown source
on a more leisurely 10 million year time scale, Proxima b's accretion would not only be too rapid for water to condense, 
 it would be too rapid for a magma surface to freeze solid unless there were no atmosphere \citep{Lupu2014}.
 
 \subsection{Proxima b: Stellar Wind}
 
Here we ask if atmospheric erosion by Proxima's stellar wind has been important.
In the Solar System, the solar wind erodes through direct collisions (sputtering)
and through its magnetic field (ion pickup). 
The latter in particular is important for Venus and Mars.
Venus intercepts about 4000 grams of solar wind per second, estimated 
using $\dot{M}_{\odot} = 2\times 10^{-14} M_{\odot}$ yr$^{-1}$ \citep{Wood2002}.
Average quiet sun observed rates of oxygen ion escape from Venus are much smaller, about 150 g $s^{-1}$ \citep{Fedorov2011}.
Modeled O$^+$ escape rates range from 150 to 800 g $s^{-1}$ \citep{Jarvinen2009}.  
Solar wind driven escape at Mars is more efficient:
Mars intercepts about 300 grams of solar wind each second, which is comparable to the observed 
O$^+$ escape rate of 160 g $s^{-1}$ \citep{Brain2015}
 and to modeled O$^+$ escape rates of 300 -- 500 g $s^{-1}$ \citep{Lammer2003b}.
Apparently the solar wind impinging on Mars erodes roughly its own mass in martian atmosphere,
although Venus suggests that escape processes are,
among other things, sensitive to $v_{\mathrm{esc}}$.
 
Suppose that the observed martian regime approaches the asymptotic efficiency.
This may be a reasonable bound on a process that in broad brush is a problem of turbulent mixing. 
If so, the planet's mass loss rate would be equal to the mass of stellar wind intercepted,
\begin{equation}
\dot{M} = \frac{\pi R^2 }{ 4\pi a^2} \dot{M}_{\mathrm{sw}} ,
\end{equation}
where $\dot{M}_{\mathrm{sw} }$ is the star's mass loss rate.
The stellar wind is assumed strong enough that the planet's cross section to the wind is comparable to its physical cross section.
Proxima b therefore intercepts $1\times 10^{-6}$ of $M_{\mathrm{sw} }$.
We estimate that $\dot{M}_{\mathrm{sw} }$ for Proxima today is roughly $3\times 10^{-15} M_{\odot}$ yr$^{-1}$,
in which we have scaled the solar wind by XUV \citep{Wood2002}.
If we assume that $\dot{M}_{\mathrm{sw} } \propto t^{-1}$, we estimate that Proxima b has intercepted roughly
$1\times 10^{23}$ g of stellar wind since it first became potentially habitable at 150 Myr of age.
This corresponds to $1\times 10^{-5}$ of Proxima b's mass, or about 10 bars of atmosphere.
Losses would be less than 1 bar if the extrapolation were based on Venus.
These estimates are respectively 2 and 3 orders of magnitude smaller than what XUV can do in the first 150 Myr for an H$_2$-rich or H$_2$O-rich atmosphere \citep{Barnes2017}.
On the hand, the stellar wind might pose the greatest existential threat to a CO$_2$ atmosphere at times after the first 150 Myrs.

\subsection{Proxima b: ``Ashes, Ashes and Dust, and Thirst there is''}

Proxima b may have no good analog in the Solar System. 
Venus seems the best candidate. 
Venus retains very little water (in the atmosphere, $5\times 10^{-6}$ of Earth, \citet{Fegley2014}) and probably very little in its interior. 
\citet{Gillmann2009} and \citet{Hamano2013} explain Venus as having been thoroughly desiccated by hydrogen escape 
from a steam atmosphere over a molten silicate surface that lasted for more than 100 Myrs.
In this picture the mantle remained in equilibrium with the water vapor in the atmosphere, and thus the loss of all
the water from the atmosphere meant also the loss of almost all the water from the mantle.
Both early insolation and accretional energy are greater for Proxima b than for Venus,
which makes Venus's story seem all too likely Proxima b's story as well.

Mercury and Io represent end members where the forces of escape (for Mercury, insolation and XUV; for Io, impact erosion and thermal radiation from Jupiter when young) 
are almost wholly victorious.  Both retain sulfur.  Notably, Mercury appears to retain about as much water as it can harbor in its shadowed craters,
which is empirical evidence either that late high speed impacts by comets or asteroids do not wholly preclude the accretion of small amounts of water,
or that even a planet as blasted as Mercury can still degas a little water.        
Mars may be a guide to what Proxima b might look like if it were in equilibrium with
a late bombardment (post-dating the runaway greenhouse phase) of volatile-rich bodies dislodged from cold distant orbits.
Between Mercury and Mars there is a place for a habitable desert state for Proxima b \citep{Abe2011,Turbet2016}.

Europa and Ganymede are examples of planets born with too much water to lose.
We argued with respect to Figure \ref{XUV_vs_x} that over 5 Gyr it is difficult for any planet, no matter how small or strongly irradiated, to 
selectively lose more than about 0.2\% of its mass as hydrogen, so that an initial water inventory greater
than say 2\% by mass is likely not to be lost save by impact erosion.  
Both Europa and Ganymede appear to have been heated well enough during accretion that they are thoroughly differentiated,
yet each retains a lot of water (Europa, $\sim 7$\%; Ganymede, nearly 50\%). 
How a water world Proxima b might accrete is a puzzle --- perhaps a giant impact of a stray water rich planet with a local planet
could do it, or perhaps it migrated in from a colder place in what is to date a wholly conjectural planetary system --- but given the current absence of a predictive model of planet formation, nothing should be ruled out \citep{Lissauer2007, Ribas2016,Barnes2017,Coleman2017}.
 
Earth would be the best of all possible analogs, 
but the real Earth's hydrosphere is too thin to have survived Proxima's youthful luminosity and 
the onslaught of XUV radiation and flares that continues today \citep{Davenport2016}.  
 Both \citet{Ribas2016} and \citet{Barnes2017} explicitly construct habitable states by starting with just enough
 water that the loss of the hydrogen from some 1-10 oceans of water leaves an ocean or so behind after the early steam atmosphere condensed.
A decade ago, \citet{Lissauer2007} had pointed out that an Earth-like outcome in this scenario, or in any of several other scenarios he considered, requires
  ``precisely the right amount of initial water or just the right dynamics.''  I.e., Earths are unlikely outcomes for Proxima b compared
  to a vastly greater phase space of less happy outcomes.
When conceiving {\it Kepler,} \citet{Borucki1996} set out to determine the number of Earth-like planets in the cosmos by direct observation.
With all apologies to theory, this remains the path forward.

\section{Trappist 1f}

Seven Venus-sized planets were recently discovered orbiting the very faint M-star Trappist 1 \citep{Gillon2017}.
Several of these are in the habitable zone.
One of the planets, Trappist 1f, has already been well-enough characterized by transit timing
variations that it cannot be ignored.  Trappist 1f has therefore been superposed on Figures 1-4.  
Interest ensures that other planets in the system will also soon be well-characterized, but we will pass on these here,
to return to the topic in the future if merited.

Trappist 1 is described as a red dwarf of effective $T=2560\pm 60$, mass $0.08 M_{\odot}$,
and luminosity $5.24\times 10^{-4} L_{\odot}$  \citep{Gillon2017}.
A na{\"i}ve comparison to models by \citet{Burrows2001} would suggest that it could be young, $\sim 0.5$ Gyr, and still fading,
but it is more likely that the star has reached its asymptotic main sequence luminosity and is therefore older than
1 Gyr.  In either case the star would have been about $4\times$ brighter than it is now at 100 Myr and about $20\times$ brighter when 10 Myr of age (Figure \ref{Insolation}).   Trappist 1f's current orbit spent more time
in the runaway greenhouse zone than did Proxima b's.

The star's strength as a UV \citep{Bourrier2017} and X-ray \citep{Wheatley2017} source have been reported;
we will take $L_{\mathrm{xuv}}/L_{\star} \approx 3\times 10^{-4}$, a factor 3 or so less than X-ray saturation.
Even allowing for the fading of $L_{\star}$, the ratio of the cumulative XUV fluence to that from the Sun is probably no more than the current ratio of XUV fluxes. We plot Trappist 1f at the current ratio in Figures \ref{XUV} and \ref{XUV_vs_x}.  
Trappist 1f's imaginary atmosphere appears distinctly more vulnerable to XUV radiation in Figures \ref{XUV} and \ref{XUV_vs_x}
than does Proxima b's.

We consider two cases for impact erosion on Figure \ref{Impacts}:  a lower impact velocity for material that might be getting exchanged between planets ($v^2_{\mathrm{imp}} = 0.4v^2_{\mathrm{orb}} + v^2_{\mathrm{esc}}$),
and a higher impact velocity ($v^2_{\mathrm{imp}} = v^2_{\mathrm{orb}} + v^2_{\mathrm{esc}}$) for the Trappist 1 equivalent of a Kuiper Belt or Oort Cloud. 
Other things equal, Trappist 1f's position with respect to impacts resembles Proxima b's rather closely.

Trappist 1f is roughly the same size as Earth but has a distinctly lower density, reportedly just $60\pm17\%$ that of Earth \citep{Gillon2017}. It is tempting to do this with ice, in the general pattern of Europa.  
Indeed the Trappist 1 system as a whole does resemble Jupiter and its Galilean satellites, both in scale and in relative masses.
Jupiter's system is characterized by a regular pattern of densities.
Water has clearly acted as a volatile, either in the materials from which the moons formed,
 or perhaps it has migrated from the inner moons to the outer ones in response to the luminosity of young Jupiter.
But the Solar System provides another possible analog.
At Saturn the densities of the small icy satellites are seemingly random, falling in the sequence $(1.15, 1.62, 0.97, 1.49, 1.25)$ g/cm$^{-1}$ as one marches outwards from Mimas to Rhea.
This behavior is probably a consequence of collisional evolution, with little sign that water has acted as a volatile.
A similarly random pattern of densities in the Trappist 1 system would suggest that Trappist 1f is best seen as an analog of Tethys, as iron-poor as Tethys is rock-poor. 

\section{Some outliers}

Kepler 138 is an M dwarf with 3 known planets.  The planets 138c and 138d are the same size ($\sim 1.2 R_{\oplus}$)
 but of markedly different masses.  
 The masses themselves are considerably uncertain but their ratio is not: \citet{Jontof-Hutter2015}
report $M_c/M_d = 2.96^{0.44}_{-0.35}$ with 68.3\% confidence.  
The solar system pairs Mercury/Moon (iron/rock), Io/Callisto (iron-rock/rock-ice), 
Enceladus/Tethys (rock-ice/ice) all have density ratios within a factor two,
which suggests that it may be difficult to achieve a factor of 3 by accident. 
Both planets are in the runaway greenhouse zone and thus water is not condensible.
Giving 138d an H$_2$ atmosphere can help, but asking this system to conform to photo-evaporation will not be straightforward.
A possibility that may seem promising is to invoke chemistry.  If 138c were an iron-rich body, we should
expect Fe and H$_2$O to react to make an H$_2$-dominated atmosphere with a huge scale height,
while in 138d we might imagine a rocky or watery body \citep{Jontof-Hutter2015}
with a H$_2$O-CO$_2$ atmosphere and a scale height ten times smaller.
 Under these conditions the atmosphere of the denser planet would be more vulnerable to escape. 

The planets Kepler 36b and 36c present a similar challenge. 
The two planets are in closely adjacent orbits circling an evolving (brightening) subgiant star.
The outer planet (c) is about 70\% more massive
and almost 300\% more voluminous.  The densities differ by a factor of ten.
To first approximation the smaller one is a super-Earth and the bigger one is a sub-Saturn or a gassy Neptune.
\citet{Owen2016b} devised a photo-evaporation story in which the two planets were initially of similar composition,
but the inner one being less massive from the start evaporates first, and they showed that this could be made to work. 
Curiously, it is now the smaller of the planets that has the bigger escape velocity and the deeper potential well,
which probably requires that the interior of the inner planet was much hotter during escape than it is now.  
But of course it is also possible that they were just made this way.

All 3 known planets of Kepler 51 (b,c,d) are reported to be of exceedingly low density \citep{Masuda2014}.
They are outliers in Figures 1-4.
These planets as described are not really stable, and thus there is cause for some skepticism of their reported attributes  \citep{Cubillos2017}.
\citet{Masuda2014} reports a considerable uncertainty in the diameter of the star.
It is possible that the star has been misunderstood.

\section{A Discussion of Isothermal Escape}
\label{discussion}

The isothermal approximation is (i) simple; (ii) consistent with $\sigma T^4$ radiative cooling; 
(iii) equally applicable to all planets;
and (iv) only one of several crude approximations we have made to poorly constrained factors that are important to escape,
and it is probably not the worst of them.
The isothermal approximation raises two different questions.  
One is how well does the isothermal approximation describe the wind, and the other is how
well does an isothermal wind approximate the rate of thermal escape. 

In principle the first question --- the detailed temperature structure of the planetary wind --- is answerable only through direct observations of planetary winds.
Lacking these, models can be constructed that contain more physics.
Detailed modeling of this sort is far beyond the scope of this paper.
What we can address within the scope of the paper is whether the energy required by isothermal escape
contradicts the assumption of isothermal escape. 

We have in this paper used isothermal models to estimate escape rates from small planets with condensed 
surface volatiles at one extreme, and for highly irradiated EGPs at the other extreme.
For the former, we show in detail in Appendix A that the energy needed to remove a planet's
volatiles on astronomical timescales of tens or hundreds of millions year, timescales appropriate to Figure \ref{Insolation}, 
is small compared to the energies associated with radiative heating and cooling.
Figures \ref{local234} and \ref{global_europa} of Appendix A are illustrative
examples made by moving famous dwarf planets to warmer places.
 
For the EGPs, temperatures in the upper atmosphere are universally expected to be much hotter than the underlying atmosphere 
because no effective means of cooling the gas have been identified at temperatures below $10^4$ K \citep[e.g.,][]{Koskinen2014}.
Thus escape from EGPs is often approximated as XUV-energy-limited, with a base near a homopause
set at the top of the underlying mixed atmosphere. 
Where this may fail is that, because escape of $10^4$ K hydrogen is relatively easy, the thermosphere may not be the true bottleneck to escape.  Instead, the bottleneck may be at the homopause, which is controlled by the magnitude of turbulent mixing,
which in turn is controlled by the total energy flux moving through the planet.   
The isothermal approximation by contrast includes the entire atmosphere in escape, and it involves
all the incident sunlight.  For the isothermal model, the bottleneck is at the bottom of the isothermal zone.
 The upgrade to the isothermal model would be a more supple stratospheric 
temperature profile generated by an accurate treatment of radiative transfer, which of course depends
on the detailed chemical and particle composition of the atmosphere and the radiative properties of the hot gases
and particles at the temperatures and pressures of the stratosphere; we will not go there. 
It is a curious coincidence that the two end-member models --- the energy-limited flux and isothermal escape ---
predict essentially the same thing when applied to planetary evaporation (compare Figures \ref{Insolation} and \ref{XUV}), but it should be noted that both models depend heavily on tidal forces to truncate the atmosphere, so that it may be that tidal truncation is the true control.  If so, a more accurate description of the tidal potential would be the direction to take further research.

The second consideration --- how well an isothermal wind approximates the rate of escape ---
is addressed in Appendix B by comparing isothermal escape to escape predicted
by some tractable alternative temperature structures for the exemplary small ocean world. 
For this purpose we construct polytropic planetary winds that are both hotter and colder than the isothermal wind;
 we construct planetary winds that are vapor-saturated at all heights;
and we add a heavy ballast gas to weigh down the atmosphere.
The saturated atmosphere, which takes it energy from the latent heat of condensation,
 is meant to set a lower bound on the temperature of the wind.
 We have described this case elsewhere \citep{Lehmer2017}, but we include full documentation in Appendix B for completeness.
The several models are compared in Figures \ref{one}, \ref{three}, and \ref{ballast2}.
For the small worlds, the uncertainty stemming from the isothermal approximation is asymmetric, 
with the isothermal approximation more likely to underestimate escape than overestimate it.
The biggest source of uncertainty is the molecular weight of the escaping gas.
Expressed in terms of the cosmic shoreline,
isothermal escape might overestimate $v_{\mathrm{esc}}$ by $15\%$ but might underestimate it by $30\%$ or more. 
Uncertainty in the insolation $I$ is larger but more symmetric, of the order of $\pm 60\%$.
The overall uncertainty stemming from the isothermal approximation
in the placement of the shoreline is probably no larger than other uncertainties in Figure \ref{Insolation}, 
such as the sizes and masses of the planets.

\section{Conclusions}
 
In this paper we have discussed the empirical evidence for a cosmic shoreline uniting the worlds of the Solar System with the exoplanets
whilst dividing the worlds between those with apparent atmospheres and those without. 
This is done through four figures, each of which compares the potency of a loss process to the planet's ability to hold an atmosphere,
and each of which shows roughly the same pattern. 
In a general sense, we have approached the problem of planets as a question of nature vs.\ nurture.
We have tried to make the best case for nurture as a determining cause.  Our bias in this direction is based on the 
fact that volatiles like H$_2$O are the most abundant condensible substances in the cosmos.
 It is therefore reasonable to hypothesize that the evolution of planetary volatiles is, in general, a story dominated by volatile loss,
 and that volatile loss provides, in general, an arrow in time.
The overall pattern of planets provides some support for this, but the case is not overwhelmingly strong.
There are too many oddities like Kepler 138c and d.
 We might argue that the case for nature is weaker still, but this would only be true if we downplayed the role of
 chance in shaping planets.  In truth, chance must play a huge role in how planets form and evolve, and
 chance does not fit well within the bounds of either nature or nurture.

Figure \ref{Insolation} finds the known worlds sorting themselves according 
to a simple $I \propto v^4_{\mathrm{esc}}$ power law relating total insolation to escape velocity. 
 Thermal escape driven by the total insolation is best regarded as a function of the sound speed $c^2_{\circ}=k_BT/m$,
because it is as sensitive to the mean molecular weight $m$ as it is to the temperature $T$. 
It is reasonable then to expect the shoreline to look like $c^2_{\circ} \propto v^2_{\mathrm{esc}}$.
We expect that $m$ will be of order 20 for terrestrial planets, $\sim \! 2.4$ for 
the cooler giant planets, and of order 1 for the hottest planets in which H$_2$ is dissociated to atoms.
If by Stefan-Boltzmann's law $I \propto T^4$, and if by chemistry we ask that $m \propto T^{-1}$,
we can recover the empirical $I \propto v^4_{\mathrm{esc}}$ relation.
We also show that a relatively simple thermal evaporation model with tidal truncation provides a credible
boundary to the highly irradiated extrasolar giant planets (EGPs).
Total insolation provides more scope for explaining the enhanced erosion 
seen in the most massive and most strongly heated EGPs than does energy-limited XUV-driven
escape, which by construction depends only on XUV irradiation and does not depend on $T$ or $m$. 
In particular, total insolation-driven escape can greatly exceed the XUV-driven energy limit.   

Figures \ref{XUV} and \ref{XUV_vs_x} address energy-limited XUV-driven escape.
The relevant quantity --- the cumulative historic XUV
irradiation at each planet, which is dominated by the excesses of the young star --- is not an observable.
 Both plots are constructed by creating a proxy quantity $I_{\mathrm{xuv}}$ that is scaled from the Sun;
 the plots are then to be regarded as comparing other systems to the Solar System.
 The plots can also be viewed as proxies for stellar wind-driven escape, as stellar winds
 stem from the same sources as the nonthermal X-rays and EUV radiations. 
 Figure \ref{XUV} looks much like Figure \ref{Insolation} because the Sun is the same in a relative way on both plots.
 The figures differ only in how the exoplanets are plotted. 
 Here we see a wider scatter of small planets that according to our hypothesis would have to be airless,
 and we see a tidier distribution of highly irradiated EGPs.   
 We confirm that energy-limited XUV-driven escape also provides a credible quantitative boundary to the EGPs.

 Figure \ref{XUV_vs_x} explores XUV-driven escape more generally in terms of a
 scaling parameter that encapsulates the $I\propto v_{\mathrm{esc}}^3\sqrt{\rho}$
 shorelines predicted by energy-limited escape.
 The figure provides context in which to discuss the limits to diffusion-limited flux.
 The division between planets that are born with too much hydrogen to lose corresponds to about 
  0.2\% H$_2$ by mass or, equivalently, about 2\% H$_2$O by mass.
  \citet{Owen2013} previously reached a similar conclusion based on energy considerations.
 The diffusion bound is an upper bound because it does not taken into account that the energy-limited flux is often the smaller,
 as it would have been for Earth itself 
 (whose upper bound on selective H$_2$ escape is less than 0.05\% by mass),
 and it does not account for escape slowing down as the atmospheric mixing ratio of H$_2$ shrinks.

Figure \ref{Impacts} addresses the competition between impact delivery of volatiles
and impact erosion of atmospheres as an alternative to irradiation-driven escape.
Here we expect the shoreline to take a simple form in which the typical impact velocity $v_{\mathrm{imp}}$
is proportional to the escape velocity $v_{\mathrm{esc}}$.  
Unfortunately impacts are poorly described by a single size of a single composition striking at a single impact velocity.
Rather, stray bodies have many sources, so that even
in our Solar System there is considerable uncertainty in the cumulative effects produced by the different sizes,
compositions, velocities, and impact geometries of the impacting bodies.
What we have done for the extrasolar planets is equate $v_{\mathrm{imp}}$ to the circular
orbital velocities $v_{\mathrm{orb}}$ of the planets, because for impact velocities to be high enough to matter, the velocities of the 
colliding bodies will be determined by the gravity of the central star.
The empirical impact shoreline then follows the quantitative relation 
 $v_{\mathrm{imp}} \approx 4\!-\!5\, v_{\mathrm{esc}}$. 
  The proportionality constant agrees with what one would extrapolate from 
  the observed consequences of the impact of Comet Shoemaker-Levy 9 with Jupiter.
  In the Shoemaker-Levy 9 impacts the bulk of the ejecta, much of which was shocked jovian air as fingerprinted by
  the chemical composition of the ejecta, were launched
  at 12-15 km s$^{-1}$; i.e., at 20-25\% of the impact velocity \citep{Zahnle1996}.
  This gives one some reason to think that the factor of $4\!-\!5$ might be generally relevant to modest impacts in deep giant planet atmospheres.
 It may be reasonable to expect a lower threshold for impact erosion from planets with 
 well-defined surfaces \citep{Melosh1989}, yet the 
 different fates of Titan and Callisto can be nicely accounted for by the same factor of $4\!-\!5$.  
 
 Finally, it has probably not escaped the reader's attention that throughout this essay we have conflated giant planets
 (with giant atmospheres) with the Solar System's terrestrial planets (with thin atmospheres).
 This is partly necessity ---  we do not yet have the tools to identify thin atmospheres amongst the exoplanets ---
 but there is also philosophy. 
 First, there is interest: the cosmic shoreline is where we live and is where we think life is likeliest to flourish. 
 Second, we do not know if the shoreline is broad or narrow (i.e., whether the transition from a thin atmosphere
 to one too thick and deep to be habitable to an ecology like our own is gentle or abrupt), 
 nor in what ways our Solar System is representative or unrepresentative of extrasolar systems.  
 In this essay we have chosen to arrange things to unify the many worlds of the cosmos.

\section{Acknowledgements}
All credit should go to the discoverers of the exoplanets for the new worlds that they have given us.  
The authors thank in particular CZ Goldblatt, MS Marley, and VS Meadows for pestering them for 9 years to write this up.
We also thank E Agol, N Batalha, P Cubillos, E Kite, JJ Lissauer, and C Johnstone for useful comments on the manuscript.  
This work was funded by NASA Planetary Atmospheres grant NNX14AJ45G and NASA
Astrobiology Institute's Virtual Planetary Laboratory under Cooperative Agreement Number NNA13AA93A.

\newpage
\appendix

We have used the isothermal approximation as our base model for thermal escape
because (i) it is simple; (ii) it is self-consistent with $\sigma T^4$ radiative cooling; 
(iii) it can be applied to all planets;
and (iv) it is only one of several crude approximations we have made to poorly constrained factors that are important to escape,
and probably not the worst of them.
An incomplete list of other doubtful but important approximations includes (i) treating a synchronously-rotating planet as a global average; (ii) treating the tidal potential as spherically symmetric; (iii) inventing stellar XUV radiation histories;
and (iv) treating the entries in a particular catalog of exoplanets on August 26, 2016 as true.  
Nevertheless, we have made a case here for purely thermal escape, as opposed to purely XUV-driven escape, by
exploiting the simplicity of the isothermal approximation.  Here we address some of the limitations
of that approximation.

\medskip
Application of the isothermal approximation raises questions that fall into two distinct categories. 
One is the degree to which an isothermal state is representative of real planetary winds. 
In principle this requires either exact solutions or observed examples.
Having neither in hand, we will address instead the energy balance of the atmosphere
and the narrower question of whether isothermal escape places excessive demands on the energy budget.. 
The second consideration is the extent to which deviation from the isothermal state is consequential to escape. 
This is addressed by comparing escape predicted by the isothermal approximation to escape predicted
by some tractable alternative temperature structures.

\section{Energy Budgets}

In general, energy must be added to an escaping atmosphere if it is to remain isothermal. 
This is because the hydrodynamic wind is driven by pressure, and thus the expanding atmosphere does work and would
cool adiabatically in the absence of heating. 
The question to be addressed here is whether escape perturbs the temperature
by enough to contradict the isothermal assumption.

The isothermal wind is described in section \ref{section:CC}
of the main text.
First, we point out that the actual ``isothermal'' assumption is that $c^2_{\circ}$ is a constant of the atmosphere; i.e., that $T/m$ be constant.
The ``isothermal'' atmosphere is not in fact isothermal if the molecular mass $m$ changes.
It is easy to imagine atmospheres in which photochemical processes break larger molecules 
into smaller ones.  For example, H$_2$O ($m=18\, m_{\mathrm{H}}$) might break down to 
${\mathrm{H}}_2 + {0.5}{\mathrm{O}}_2$ ($m=12\, m_{\mathrm{H}}$) or to
$2\mathrm{H} + \mathrm{O}$ ($m=6\, m_{\mathrm{H}}$).  
Thus the temperature at the 
critical point could be a third of what it is at the surface,
 yet the atmosphere remain well-described by constant $c^2_{\circ}$.  

The required heating to maintain an isothermal wind is obtained from conservation of energy.
For a single component fluid in steady-state in 1-D spherical symmetry, conservation of energy can be written
\begin{equation}
\label{energy}
\frac{u}{\gamma-1}\frac{\partial p}{\partial r} + 
\frac{\gamma }{\gamma-1} {p\over r^2}\frac{\partial (ur^2)}{\partial r\phantom{u^2}}  = 
 {1\over r^2} {\partial\over \partial r}\!\!\left(\! r^2k {\partial T\over \partial r} \right) + 
 \Gamma_{\mathrm{h}} - \Gamma_{\mathrm{c}} .
\end{equation} 
where $k$ is the thermal conductivity.
The parameter $\gamma$ is the usual ratio of specific heats.
Radiative heating and cooling are represented by
$\Gamma_{\mathrm{h}}$ and $\Gamma_{\mathrm{c}}$, respectively.

\subsection{The local energy budget of an isothermal planetary wind}

With $T$ constant, Eq \ref{energy} reduces to  
\begin{equation}
\label{local}
\frac{\phi r_s^2 c_{\circ}^2}{r^2} \left({2\over r} + {1\over u} {\partial u \over \partial r} \right) = 
 \Gamma_{\mathrm{h}} - \Gamma_{\mathrm{c}} .
\end{equation} 
which we have written in terms of the flow velocity $u$ and the constant mass flux $\phi=\rho u r^2$.
The velocity gradient $(1/u)(\partial u / \partial r)$ for the isothermal wind is given by Eq \ref{parker} of the main text. 
Equation \ref{local} allows us to compare the energy spent on escape to radiative heating at all heights.

The most important heating elements are stellar UV and XUV radiation and thermal
radiation from the planet's surface or lower atmosphere. 
The importance of UV heating of H$_2$-rich planetary atmospheres has been widely discussed in the literature
since \citet{Urey1952}.  
 
Planetary thermal radiation can also be important if the atmosphere contains IR-active molecules such as H$_2$O.
In the plane-parallel gray approximation, thermal radiation tends to drive the upper atmosphere
to an asymptotic isothermal state with $T_{\infty} = T_{\mathrm{eff}}/2^{0.25}$.  
In reality, if planetary thermal radiation is the major driver,
the temperature declines monotonically with increasing altitude,
in part because a declining temperature is a general property of a non-gray atmosphere \citep{Marley2014},
and in part because we are considering very extended atmospheres.

We will {\it not} here attempt to compute the radiative transfer or to solve the atmosphere's temperature structure
self-consistently; our present purpose is to determine how greatly escape affects the energy budget.
Thus we wish to estimate the magnitudes of the chief sources of diabatic heating ($\Gamma_{\mathrm{h}}$)
and compare these to the energy required to maintain isothermal escape. 

\subsubsection{alt-Europa}

  \begin{figure}[!htb] 
   \centering
   \includegraphics[width=1.0\textwidth]{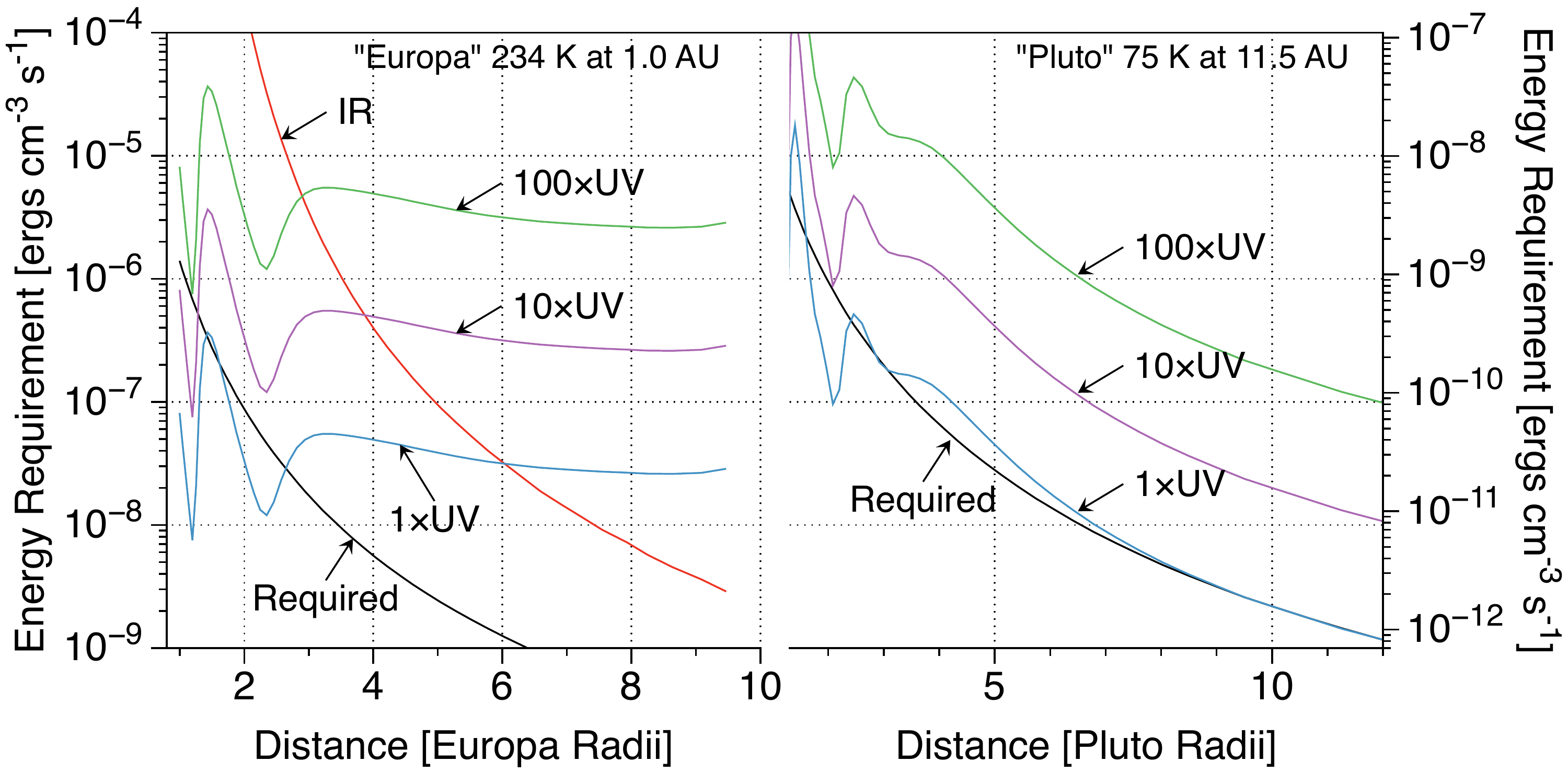} 
   \caption{\small  {\it Left.} The energy required to maintain an isothermal state in a water vapor
   atmosphere escaping from a warm Europa, 
   as a function of height (black curve), compared to the available energy sources.  
    Isothermal Europa loses its ocean in 30 Myrs.
   The total FUV+XUV radiation absorbed is shown at 3 levels of solar activity, with $1\times$ being the modern Sun.
   The curve labeled IR is thermal emission from the surface absorbed by water vapor in the low pressure limit.
   {\it Right.} The energy required by an isothermal atmosphere escaping from a warm Pluto
 (black curve). 
    The atmosphere is predominantly N$_2$, with 1\% CH$_4$ and 10 ppmv of C$_2$H$_2$.
    Alternative Pluto loses a volatile inventory equal to 1\% of its mass in 100 Myrs.
    For the modern Sun,  isothermal escape is almost exactly balanced by XUV+FUV heating.
    Higher levels of UV would make the atmosphere hotter, or would need to be balanced by radiative cooling
    if the isothermal approximation is to hold.
      }
\label{local234}
\label{pluto}
\end{figure}

We consider two specific examples inspired by cases discussed in the text.
Figure \ref{local234} explores the isothermal energy budget of an evaporating Europa.
This alternative Europa is identical to the real Europa in its mass, girth, and ocean,
but it has been placed at 1.0 AU for easier access by a lander.
The albedo is 50\% and the ice has a temperature of 234 K. 
Alternative Europa will lose its ocean in 30 Myrs.
The atmosphere is assumed to be water vapor in saturation vapor pressure equilibrium with the surface.
There is no photochemistry, a topic that we will address 
in concert with the differential escape of oxygen and hydrogen in future work.
Heating is by absorption of solar XUV and FUV radiation
at three levels of solar activity, spanning the range from the modern Sun to its active youth.  
The solar spectrum is approximated by 16 wavelength bins from $0$ to $200$ nm.
Optical depth at each wavelength is approximated by the vertical column.
The magnitude of the terms involved in the radiative transfer of
thermal infrared radiation emitted by the surface is approximated for H$_2$O
using a modern line list computed for 230 K in the low pressure limit
(Richard S.\ Freedman, pers.\ comm.). 
We assume a 234 K black body source that near the surface subtends a half-space.
The absorption there is $4.3\times 10^{-15}$ ergs s$^{-1}$ molecule$^{-1}$.
Because the atmosphere is greatly distended, it is important to take into account that the solid angle subtended
by the planet becomes much smaller at higher altitudes. 

It is apparent in Figure \ref{local234} that the temperature structure of this atmosphere will be dominated
by radiative heating and cooling.  Escape is only a minor perturbation.  
To the extent that the temperature is determined by planetary thermal radiation,
we would expect the atmosphere to cool with height.  
But at higher altitudes the heating is dominated by solar XUV and FUV, and the temperature
should rise again.   
We have not included radiative transfer nor non-LTE cooling; 
   our purpose here is only to show that the energy being
   exchanged by radiative transfer much exceeds the energy being expended driving escape.
   The bottom line is that escape on a 30 Myr time scale should result in only a modest perturbation 
   on the temperature at all heights.  
   The real atmosphere may or may not approach an isothermal state, but escape will not
   be the chief factor deciding this.

\subsubsection{alt-Pluto}

Our second example is an alternative Pluto.
The active photochemistry of CH$_4$-N$_2$ atmospheres of Pluto {\it et al} means that
many different molecules will be present to interact with light.
We set the CH$_4$ abundance to 1\% that of N$_2$, comparable to their relative volatilities,
and add additional opacity of $10^{-19}$ cm$^2$ per molecule between 160-220 nm
from other organics present at the 10 ppmv level;
the cross-section, wavelength range, and abundance approximate those of acetylene \citep{Benilan2000}. 

The Pluto in Figure \ref{pluto} was chosen so that escape would be fast enough to remove
a volatile inventory equal to 1\% of Pluto's mass in 100 Myr.  This meets the standard of rapid escape on astronomical
time scales.  The surface is 75 K, with deep nitrogen oceans partially covered by 
 floating methane polar caps, all under a partly cloudy orange sky. 
 This Pluto would be 11.5 AU from the Sun if its albedo were 30\%.
The atmosphere is enormous, extending beyond fifteen Pluto radii.
Escape is just fast enough that the energy budget of the isothermal breeze is almost exactly balanced by 
solar XUV+FUV heating at modern levels.  A more active ancient Sun would either raise
the temperature of the atmosphere, raise the rate of escape, or be balanced by radiative cooling.
Here we have not attempted to build a complete or self-consistent model.
Our point here is that energy considerations do not preclude the isothermal approximation even in this
apparently unfavorable case.

\subsection{The global energy budget of a planetary wind}

  \begin{figure}[!htb] 
   \centering
   \includegraphics[width=0.8\textwidth]{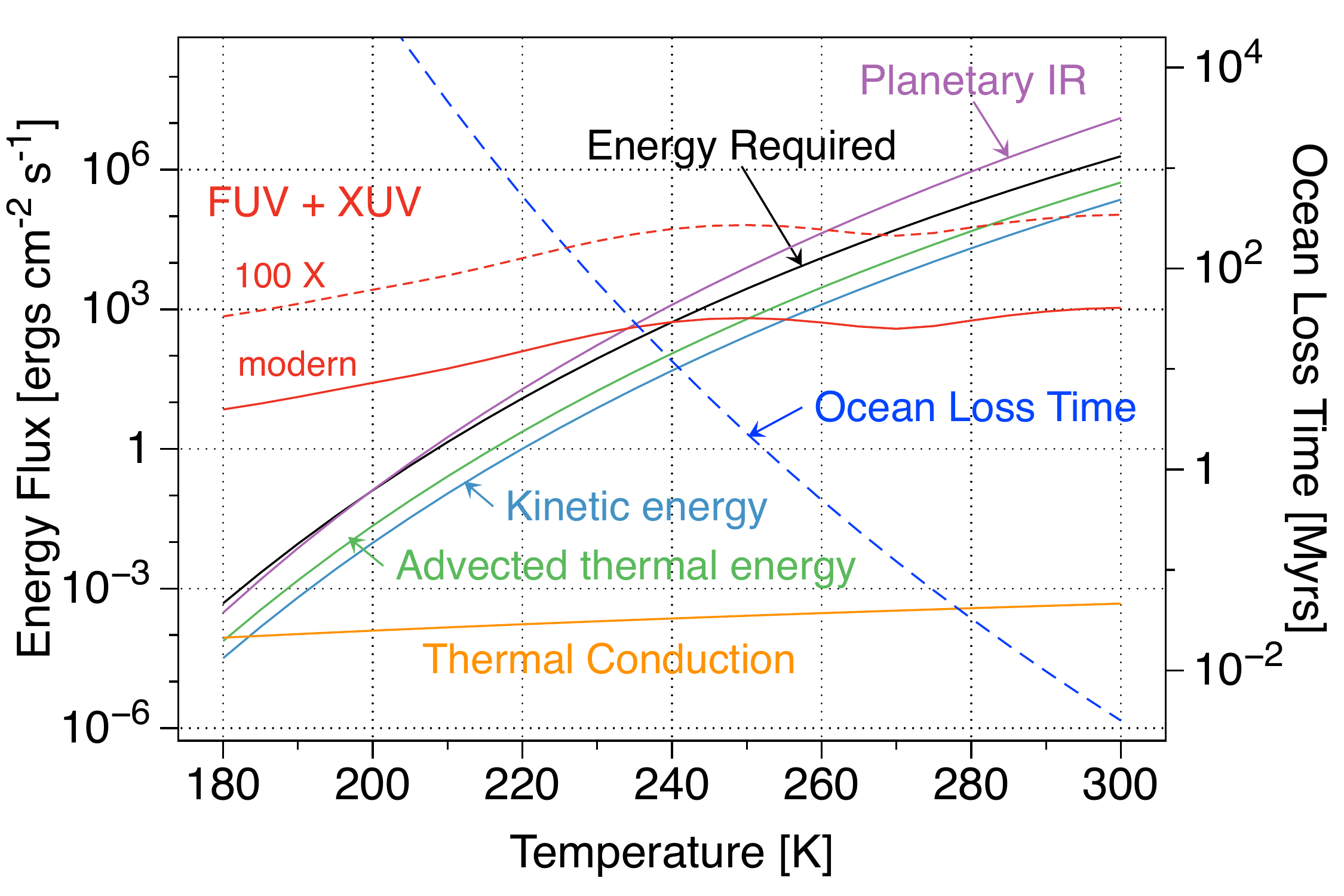} 
   \caption{\small The global energy required by escape 
   (solid black curve), compared to some other energy terms pertinent to escape.  
    The example depicts a continuum of evaporating Europas with different surface temperatures.
    How long it takes for each Europa to lose its ocean if escape is isothermal is indicated by the black dashed curve. 
   The total FUV+XUV radiation absorbed is computed using the isothermal water vapor atmosphere for $\rho(r)$.
   The stellar FUV+XUV heating is large in part because the absorbing volume of the atmosphere is vastly bigger than Europa itself.
   The curves labeled ``advected thermal energy'' and ``thermal conduction'' are indicative of the order of magnitude 
   of these terms in non-isothermal atmospheres.
   Thermal conduction is shown, but it is negligible because the distances are large and the temperature gradients are small.
   The curve labeled IR is a lower bound on the absorption of thermal emission from the surface.
   No radiative transfer is performed.  The purpose here is merely to show that the energy being
   exchanged by radiative transfer exceeds the energy spent boosting gas into space. }
\label{global_europa}
\end{figure}

A different way to look at energy is through the global energy budget obtained by integrating Eq \ref{energy} over the whole atmosphere, from a lower boundary at $r_s$ to the critical distance $r_c$.
For a one-component fluid with constant $m$ and constant mass flux $\phi$, the integral is
\begin{equation}
\label{global}
\phi \left( {u_c^2\over 2} - {u_s^2\over 2} + {\gamma\over \gamma -1}\left({p_c\over \rho_c} - {p_s\over \rho_s}\right) -{GM\over r_c} + {GM\over r_s} - k_c\left({\partial T\over \partial r}\right)_{\!\!c} + k_s\left({\partial T\over \partial r}\right)_{\!\!s} \right) = \int^{r_c}_{r_s} r^2dr\left(\Gamma_{\mathrm{h}} - \Gamma_{\mathrm{c}}\right) .
\end{equation} 
Note that the energy required to evaporate a volatile from the condensed phase is not included in the
energy budget of the atmosphere. 

Equation \ref{global} is a statement of conservation of energy for the escaping 
atmosphere as a whole. 
 Setting the upper bound at $r_c$ is appropriate to the
 transonic wind because nothing that happens above $r_c$ can be communicated to $r_s$.
 This includes any XUV heating.  This restriction does not apply to subsonic winds,
 for which the energy balance needs to account explicitly for the upper boundary condition that prevents
 the wind from freely expanding, as well as any radiative heating or cooling that takes place before the wind reaches the upper boundary.  
On the left hand side, Eq \ref{global} accounts for the net gains of kinetic, thermal, and potential energy, and the flow of heat into or out of the volume by thermal conduction.  The right hand side includes radiative heating and cooling, and 
it implicitly includes other heating terms (e.g., breaking waves, friction etc.) that may be pertinent.
If there are chemical changes induced by photolysis, the $\gamma$ would differ at the top and bottom, and
the net endothermicity of the chemical changes needs to be taken into account.
This is usually accounted for as a component of an empirical ``heating efficiency.'' 

Figure \ref{global_europa} depicts a continuum of evaporating Europas as a function of surface temperature.
For illustration, 
we estimate the order of magnitude of the thermal energy term by setting $\gamma=1.4$ (the room temperature value
for a diatomic molecule) and assuming that $p_s/\rho_s-p_c/\rho_c \approx p_s/\rho_s$; we are agnostic with respect to the sign.  
 The approximate magnitude of thermal conduction is estimated by setting the temperature gradient equal to
 $T_c/(r_c-r_p)$.  
 The thermal conductivity of water vapor is approximated by extrapolating a fit to 
 values measured between 300 and 600 K, 
$k(\mathrm{H}_2\mathrm{O}) = 1.1\, T^{1.3}$ ergs cm$^{-1}$K$^{-1}$sec$^{-1}$. 
 It is generally true that thermal conduction can only be important
 over distances that are very small compared to the extended atmospheres considered here,
 and thus thermal conduction is not a major player when escape is taking place.   
 
 The Europas of Figure \ref{global_europa} divide naturally into cold long-lived oceans with energy
 budgets dominated by solar FUV and XUV heating, and warm short-lived oceans
 with energy budgets dominated by planetary thermal radiation and escape.
 The latter can be conventionally habitable with liquid water at the surface, but they do not last long.
 For the colder planets, the relatively high levels of FUV and XUV heating suggest that
  the isothermal approximation probably underestimates the temperature
 and the rate of escape, while for the warmer planets the energy demands set by rapid escape
 suggests that the isothermal approximation overestimates
 the temperature and the rate of escape. 

\section{Alternatives to isothermal escape}

Here we put the limitations of the isothermal approximation into broader context.
We compare escape rates predicted by the isothermal model to (i) escape rates predicted by polytropic atmospheres;
(ii) escape rates predicted by a fully vapor saturated atmosphere;
and (iii) the beneficial consequences of adding a heavy ballast gas.
Another consideration is changes in molecular weight caused by photochemistry.
All of these considerations, taken together, provide some guidance to the breadth of outcomes that might be expected
in mapping the cosmic shoreline against $v_{\mathrm{esc}}$ or insolation $I$.  

\subsection{Polytropes}

The polytropic equation of state can be defined by  
 \begin{equation}
 \label{polywind0}
 p \propto \rho^{n+1} ,
  \end{equation}
where $n$ is the polytropic index.
 For an ideal gas with constant mean molecular mass $m$, the temperature goes as 
 \begin{equation}
 \label{polywind1}
T \propto \rho^{n} ,
  \end{equation}
 although strictly it is the ratio $T/m$ that goes as $\rho^{n} $.
The polytrope generalizes the adiabatic relation $p \propto \rho^{\gamma}$ 
to values of $n$ that can be positive or negative.
The formal relation between $\gamma$ and the polytropic index $n$ is $\gamma=n+1$. 

Equations \ref{polywind1} and Eqs \ref{continuity} and \ref{force} of the main text 
can be combined into a
planetary wind equation that for the polytrope looks superficially like the isothermal wind of Eq \ref{parker},
differing only by a factor $n+1$ in the sound speed,
 \begin{equation}
 \label{polywind2}
 \left( u^2 - \frac{(n+1) p}{\rho} \right)\frac{1}{u}\frac{\partial u}{\partial r}
  = \frac{2(n+1) p}{\rho r} - \frac{GM}{r^2} .
 \end{equation}
 The critical point conditions for the transonic solution also look like those of the isothermal wind,
 \begin{eqnarray}
 \label{polywind3}
 u_c^2 &\, =  \,& \frac{(n+1) p_c}{\rho_c}  \nonumber \\
\frac{2(n+1) p_c}{\rho_c} &\,  = \, & \frac{GM}{r_c} .
\end{eqnarray}
However, unlike in the isothermal wind, the sound speed is not a constant.
Equation \ref{polywind2} can be integrated analytically \citep{Parker1963,Zahnle1986} in spherical geometry
 \begin{equation}
 \label{polywind25}
{1\over 2}{u^2\over c_s^2}  - {GM\over c_s^2r} + {n+1\over n}{T\over T_s} = {1\over 2}{u_s^2\over c_s^2} - {GM\over c_s^2r_s} + {n+1\over n} .
\end{equation}  
Here, the constant of integration has been evaluated at the surface. It is also evaluated at the critical point, and
the equations combined to obtain a transcendental expression for the velocity $u_s$ at the surface as a function of $n$ and the other properties at the surface,  
 \begin{equation}
 \label{polywind26}
{1\over 2}{u_s^2\over c_s^2}  + {n+1\over n} - {GM\over c_s^2r_s} =
 {1\over \beta}\left( {4c_s^2 r_s \over GM} \right)^{4\beta} \left( {n+1 \over 2} \right)^{(2\beta/n)}
 \left( {1\over 2}{u_s^2\over c_s^2} \right)^{\beta}
\end{equation}  
with 
 \begin{equation}
 \label{polywind27}
\beta = {n\over 2-3n} .
\end{equation}  
As in the isothermal wind, the solution for $u(r)$ is independent of density or flux.
If the surface density $\rho_s$ is known independently, 
the escape flux is $\phi(r) = \rho_s u_s r_s^2/r^2$.
 
It is often easier to obtain the desired transonic solution numerically.
 The best approach is to obtain the slope $(du/dr)_c$ at the critical point by writing it as a ratio
 \begin{equation}
 \label{polywind5}
\frac{1}{u}\frac{\partial u}{\partial r}  =  \frac{N(r,\rho,p)}{D(u,\rho,p)} ,
\end{equation}  
in which the numerator and denominator simultaneously pass through zero at the critical point,
\begin{eqnarray}
 \label{N5}
N(r,\rho,p) & \, = \, & \frac{2(n+1) p}{\rho r} - \frac{GM}{r^2} \nonumber \\
D(u,\rho,p) & \, = \, &  u^2 - \frac{(n+1) p}{\rho} .
\end{eqnarray}  
The slope at the critical point is obtained by applying L'H{\^o}pital's rule, 
 \begin{equation}
 \label{polywind6}
 \frac{N_c}{D_c} =  \frac{(dN/dr)_{c}}{(dD/dr_{c}} .
 \end{equation}  
After a bit of manipulation using the critical point conditions (Eq \ref{polywind3}) and continuity, 
we obtain a quadratic expression for $x\equiv u^{-1}_c\left({\partial  u /\partial  r}\right)_{c}$ 
 \begin{equation}
 \label{polywind7}
 \left(n+2\right) x^2  + \frac{4n}{r_c}  x + \frac{4n -2}{r_c^2} = 0 
\end{equation}  
that is easily solved to give the slope $\left( \partial u/\partial r \right)_c$ at the critical point.
(The corresponding result for the isothermal wind is $\left( \partial u/\partial r \right)_c = \pm u_c/r_c$.)
The positive root is appropriate for an expanding flow.
The negative root corresponds to accretion.
There are no accelerating planetary wind solutions with $n>1/2$. 

Equation \ref{polywind5} is integrated inwards and outwards from the critical point.
The inward integration terminates at the surface at $r_s$.
For a Clausius-Clapeyron atmosphere, the pressure $p_s$ and density $\rho_s$ at the surface $r_s$ are unique functions
of the surface temperature $T_s$.  The velocity $u_c$ is 
iterated until the desired lower boundary conditions are met.

\subsection{Saturated atmospheres of a single substance}

Next we consider the escape of an atmosphere that is in saturation vapor pressure
equilibrium at all altitudes.  We have described such atmospheres elsewhere \citep{Lehmer2017}.
The hypothesis is that such an atmosphere is the coldest and therefore least able to escape of any
atmosphere composed of a single substance.
The saturated wind takes much of the energy it needs from the latent heat of condensation. 
The caveat is that condensation must take place at all heights, a condition that may
not be met by a low density vapor as it accelerates in the vicinity of the critical point.

We again assume an ideal gas and we also assume that the usual continuity 
and force equations apply.
The polytropic relation is replaced by an equation of vapor pressure equilibrium;
i.e., saturation provides an additional relation between $p$ and $T$ that is functionally
equivalent to the polytropic relation or the isothermal assumption.
Saturation vapor pressure is approximated by the familiar 2-parameter CC relation
\begin{equation}
\label{saturation}
p = p_w e^{-T_w/T}.
\end{equation}
A very good approximation for $130\!<\!T\!<\! 270$ K over ice takes $T_w=6120$ K and $p_w =3.2\times 10^7$ bars.
The 2 parameter fit agrees with the 7 parameter approximation \citep{Fray2009} that we used for the surface pressure
of the isothermal atmosphere to better than 10\% for $T>120$ K. 
The simple 2-parameter expression for saturation vapor pressure is desirable here because we want to work with analytic 
expressions for $dT/dp$.  

Applying Eqs \ref{continuity} and \ref{force} 
to a condensing wind implicitly assumes that the
condensate has negligible mass and is also stationary, so that there there is no radial transport and hence
no net sink on the vapor. 
Saturation presumes that the increasingly unfavorable kinetics in a rapidly accelerating wind 
will not preclude the presence of condensates at all heights.
These assumptions describe a thin fog extending from the surface to
the critical point.   

It is convenient to write the perfect gas law in the same form as we used for the isothermal atmosphere
\begin{equation}
\label{perfect1}
p = \rho c^2 ,
\end{equation}
where $c$ is defined
\begin{equation}
\label{c2}
c^2 \equiv {kT\over m} .
\end{equation}
As with the polytrope, $c$ as defined by Eq \ref{c2} is not a constant.

\medskip
The goal is to combine Eqs \ref{continuity}, \ref{force}, 
\ref{saturation}, and \ref{perfect1}
 into a single planetary wind equation analogous
to Eq \ref{polywind2} that expresses the slope $\partial u/\partial r$ as a function of $r$, $u$, and $T$.
One can eliminate $p$ from Eq \ref{force} 
using Eqs \ref{perfect1} and \ref{c2}:
\begin{equation}
\label{elim}
u{\partial u \over \partial r} + {c^2\over \rho}{\partial  \rho \over \partial  r} + {c^2\over T}{\partial T \over \partial r}= -{GM \over r^2} .
\end{equation}
Saturation can be used to express ${\partial T/\partial r}$ in terms of ${\partial \rho/\partial  r}$,
\begin{equation}
\label{dTdr}
{c^2\over \rho} {\partial \rho \over \partial r} = \left( {T_w-T\over T} \right){c^2 \over T}{\partial T \over \partial r} ,
\end{equation}
and continuity eliminates ${\partial \rho / \partial r}$ in terms of ${\partial  u / \partial  r}$:
\begin{equation}
u{\partial u \over \partial r} - c^2\left( {T_w\over T_w-T}\right) \left({1\over u}{\partial u \over \partial r} + {2\over r} \right) = -{GM \over r^2}.
\end{equation}
The result is the desired planetary wind equation in the familiar Parker form of Eqs \ref{parker} 
and \ref{polywind2},
\begin{equation}
 \label{wind}
\left(u - {c^2\over u}\left( {T_w\over T_w-T}\right) \right) {\partial u \over \partial r}  = {2c^2\over r} \left( {T_w\over T_w-T}\right)  -{GM \over r^2}.
\end{equation}
Written explicitly as an expression for $\partial u /\partial r$, 
\begin{equation}
 \label{equiv}
{1\over u}{\partial u \over \partial r}  = {N(r,T)\over D(r,T,u)} ,
\end{equation}
where here the numerator $N(r,T)$ is
\begin{equation}
 \label{numerator}
N(r,T) \equiv  {2c^2\over r} \left( {T_w\over T_w-T}\right)  -{GM \over r^2} 
\end{equation}
and the denominator $D(r,T,u)$ is
\begin{equation}
 \label{denominator}
D(r,T,u) \equiv u^2 - {c^2}\left( {T_w\over T_w-T}\right)  .
\end{equation}
The critical point conditions for the transonic wind, $N_c=0$ and $D_c=0$, link $r_c$, $T_c$, and $u_c$:
\begin{eqnarray}
 \label{r_c}
r_c & \,= \,& {GMm(T_w-T_c)\over 2kT_c T_w}  \nonumber \\
u_c^2 & \,= \, &c_c^2\left( {T_w\over T_w-T_c}\right) = {GM \over 2r_c^3}.
\end{eqnarray}
The transonic solution is obtained by integating Eq \ref{equiv} numerically inward and outward from the critical point.
For the first step we need to know the slope $\left( \partial u / \partial r\right)_{\! c}$ at the critical point.
This is obtained from Eq \ref{equiv} by using L'H{\^o}pital's rule,
\begin{equation}
 \label{LHopital}
{1\over u_c}\left( {\partial u \over \partial r} \right)_{\!\! c}  = {N_c\over D_c} = { \left(dN/dr\right)_{ c} \over \left(dD/dr\right)_{ c} }.
\end{equation}
The numerator becomes
\begin{equation}
\left({dN\over dr}\right)_{\!\! c} = {2GM\over r_c^3} - {2c_c^2\over r_c^2}{T_w\over T_w-T_c} + {2c_c^2\over r_c}{T_w\over \left(T_w-T_c\right)^2} \left({\partial T \over \partial  r}\right)_{\!\! c}
+{2c_c^2 \over r_c T_c}{T_w\over T_w-T_c} \left({\partial T \over \partial  r}\right)_{\!\! c}
\end{equation}
from which $\left({\partial T /\partial r}\right)_{\! c}$ is eliminated in favor of $\left({\partial u /\partial r}\right)_{\! c}$ by using Eq \ref{dTdr},
\begin{equation}
\label{dN/dr_c}
\left({dN\over dr}\right)_{\!\! c} = {GM\over r_c^3} - {4c_c^2\over r_c^2}{ T^2_wT_c\over \left(T_w-T_c\right)^3}
- {2c_c^2\over r_c}{ T^2_wT_c\over \left(T_w-T_c\right)^3} {1\over u_c}\!\left({\partial u \over \partial r}\right)_{\!\! c} .
\end{equation}
The denominator becomes
 \begin{equation}
\left({dD\over dr}\right)_{\!\! c} = 2u_c\left({\partial u \over \partial r}\right)_{\!\! c} 
- {c^2_c\over u_cT_c}{T_w\over T_w-T_c}\left({\partial T \over \partial r}\right)_{\!\! c}
- {c_c^2\over u_c} {T_w\over \left(T_w-T_c\right)^2}\left({\partial T \over \partial r}\right)_{\!\! c}
\end{equation}
from which $\left({\partial T /\partial r}\right)_{\! c}$ is eliminated in favor of $\left({\partial u /\partial r}\right)_{\! c}$ by using Eq \ref{dTdr}
 and $u_c$ from Eq \ref{r_c}
\begin{equation}
\label{dD/dr_c}
\left({dD\over dr}\right)_{\!\! c} = \left( 2 + {T_wT_c\over \left(T_w-T_c\right)^2}\right) u_c\left({\partial u \over \partial r}\right)_{\!\! c} 
+ {2u^2_c\over r_c} {T_wT_c\over \left(T_w-T_c\right)^2} .
\end{equation}
Equation \ref{LHopital} can then be written as a quadratic equation for  $x\equiv u^{-1}_c\left({\partial u / \partial r}\right)_{c}$,
\begin{equation}
\label{quadratic2}
\left( 2 + {T_wT_c\over \left(T_w-T_c\right)^2}\right) x^2
+ {4\over r_c} {T_w T_c\over \left(T_w-T_c\right)^2}\,x
+ {4\over r_c^2}{ T_w T_c\over \left(T_w-T_c\right)^2} - {2\over r_c^2} = 0 .
\end{equation}
Again, the positive root corresponds to an accelerating flow and is the appropriate solution here.
If there is no real root to Eq \ref{quadratic2}, there is no planetary wind solution.
In these cases the atmosphere must either be static or ill-described by a fluid because 
escape occurs directly from the surface.

Comparison of Eqs \ref{polywind7} and \ref{quadratic2} reveals that the saturated wind
and the polytrope correspond exactly at the critical point for an effective value of $n$  
\begin{equation}
\label{cool}
n = {T_w T_c\over \left(T_w-T_c\right)^2} .
\end{equation}
Equation \ref{cool} is more an illustration than a useful relation, because $T_c$ must be solved
for, and $T_c$ varies from case to case.  But for the purposes of the 
illustration, with $T_c=183$, the effective value of $n$ is $0.03$. 
To put this another way, the fully vapor saturated wind --- which we suggest is as cold a wind
as a pure substance can support --- is not extremely different from the isothermal
wind for which $n=0$.

\subsection{Europa: the many worlds hypothesis}

Europa, the smallest of the Medicean planets,
 is currently of great interest to NASA as a local example of an ocean world.
Europa in its current state at its current location does not qualify as a conventionally 
habitable planet. Its surface is too cold ($T\approx 100$ K), it has no significant atmosphere,
and its liquid water is hidden under some 10 km of ice.
But in the future, when the Sun has evolved to burn helium as a so-called  ``horizontal branch'' star,
our Sun will be roughly 25 times brighter than it is today and Europa will be in the middle of the habitable zone
\citep{Nisbet2006}.  It is also possible that Europa was, for a brief moment when both it and Jupiter were young,
a warm wet world heated by thermal radiation from Jupiter itself.
 
\begin{figure}[!htb] 
   \centering
   \includegraphics[width=1.0\textwidth]{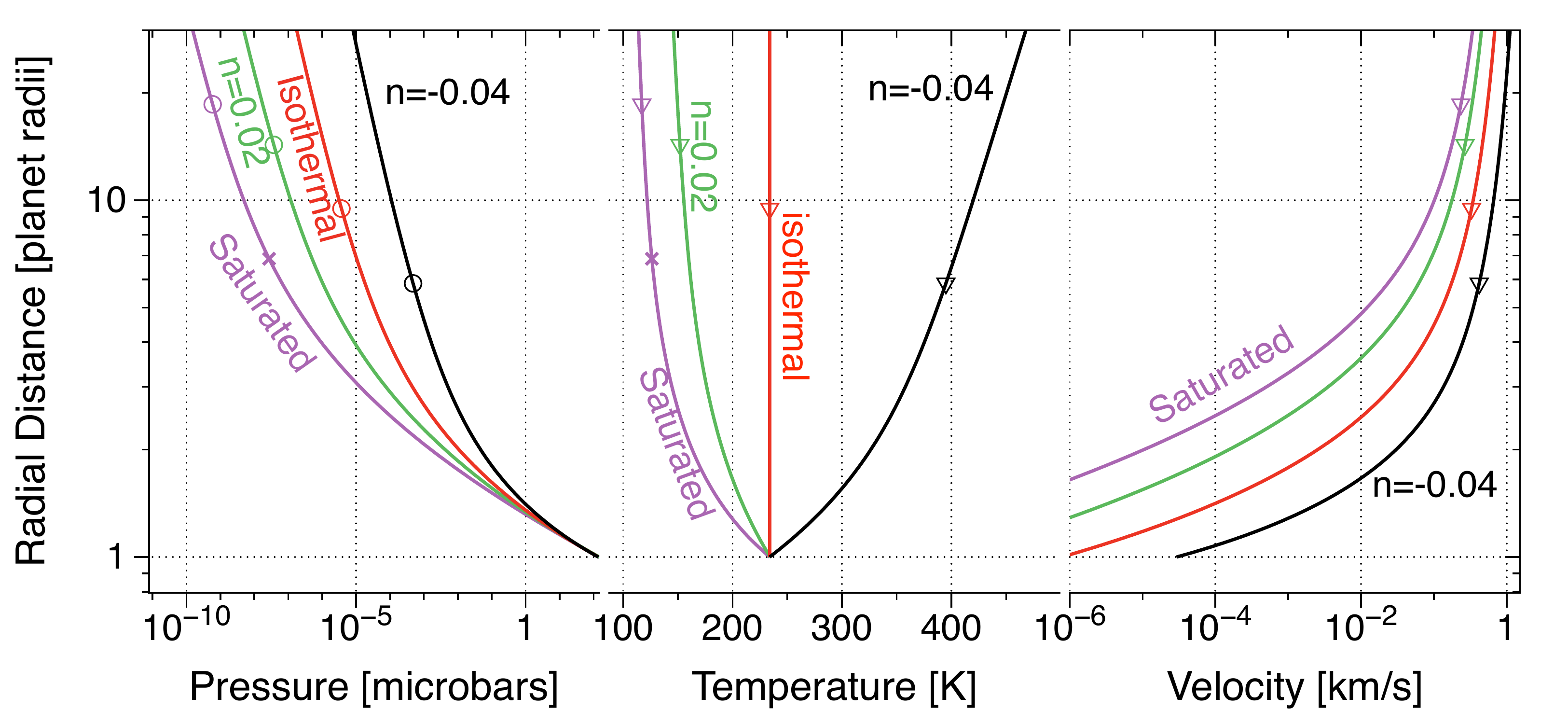} 
   \caption{\small Comparisons of vertical structures of several different transonic wind solutions
   for a water vapor atmosphere of a Europa at 1 AU from our Sun.
   The albedo is 50\%, the surface temperature is 234 K, water vapor is in saturation equilibrium with the surface.
   H$_2$O is treated as an inert gas that remains intact at all heights.
   The isothermal solution is bracketed by two polytropes, one warmer, one cooler.
   Also shown is an atmosphere that is vapor saturated at all heights.  
   The critical points are marked by open symbols and the approximate location of the exobase, if there is one, is marked by an ``x.''
   }
\label{one}
\end{figure}

Europa is very nearly the smallest world that can retain a liquid water surface for any significant
period, which makes Europa an excellent candidate for a study of the littlest habitable world.   
In Figure \ref{one} we compare four alternative atmospheres for a planet with the size and weight of Europa.
For Figure \ref{one}, Europa is placed at 1 AU from the Sun.  The Bond albedo is set to 50\%.
This makes the surface temperature 234 K.
Figure \ref{one} compares vertical structures of two polytropes and the saturated atmosphere
to the isothermal atmosphere.  All the atmospheres are very large 
compared to the planet itself.  The upper atmosphere of the $n=-0.04$ polytrope is either about twice as hot
as the surface, or its mean molecular weight is about half what it is at the surface; the polytrope
cares only about $p/\rho$.  The saturated atmosphere resembles an $n=0.03$ polytrope,
as Eq \ref{cool} suggested it might.
In this case the upper atmosphere is about half the temperature of the surface.

  \begin{figure}[!htb] 
   \centering
   \includegraphics[width=1.0\textwidth]{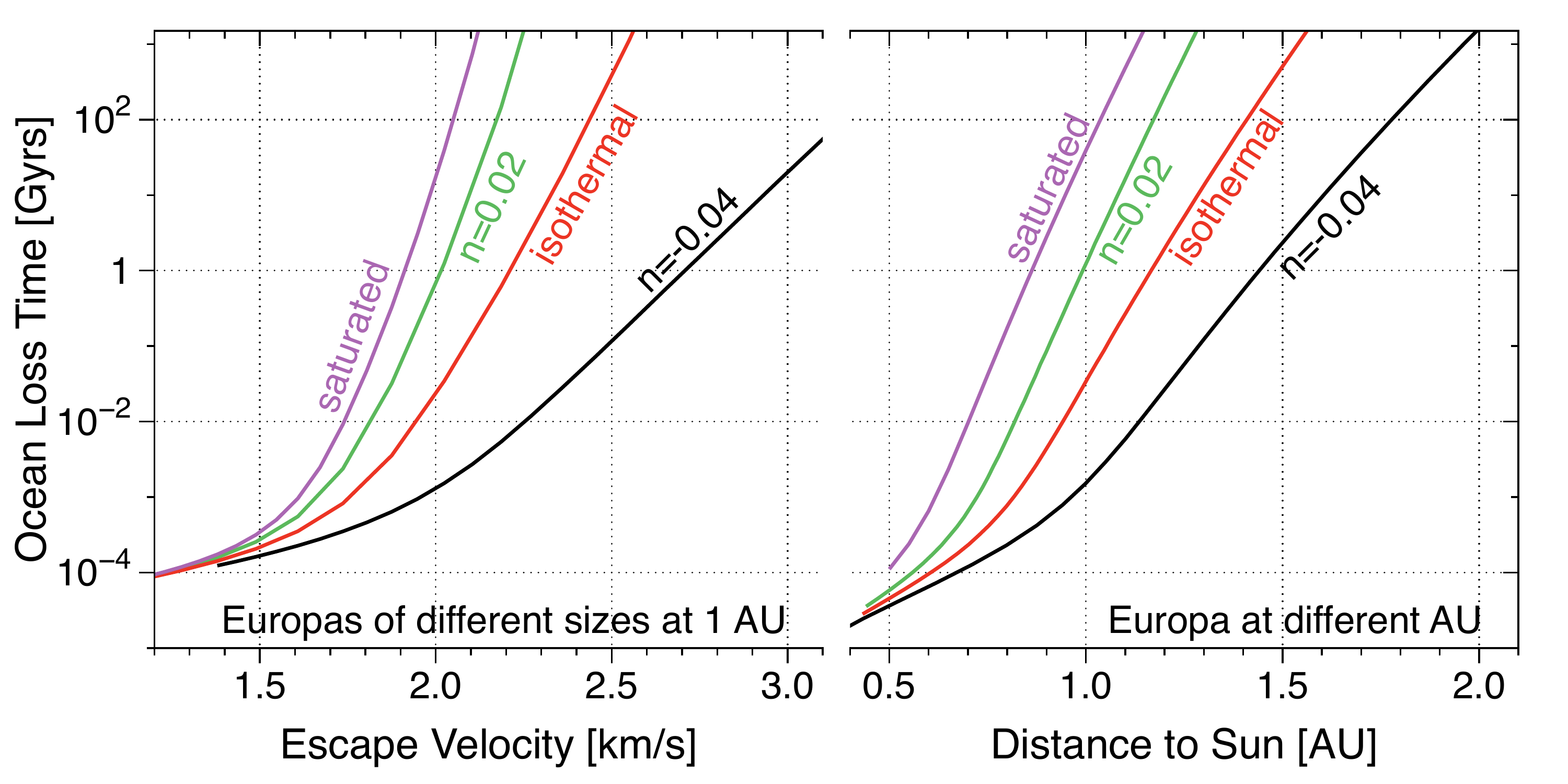} 
   \caption{\small How long it takes for alt-Europas of different sizes to lose their oceans for the four different planetary winds in Figure \ref{one}.
    The Europas in the left-hand panel are bigger or smaller than Europa itself ($v_{\mathrm{esc}}=2.0$ km s$^{-1}$.)
    The Europas in the right-hand panel are true Europas, but placed at different distances from the Sun.
     For simplicity all Europas have an albedo of 50\% even if the ocean melts. }
 \label{three}
\end{figure}
    
  Figure \ref{three} illustrates that alt-Europa's long-term survival as a conventionally habitable world is sensitive to details.
    The left-hand panel of Figure \ref{three} compares evaporation lifetimes at 1 AU predicted by the several models
     as a function of escape velocity.  
     These Europas have been scaled from the true Europa ($v_{\mathrm{esc}}=2.0$ km s$^{-1}$) at constant density,
     so that $v_{\mathrm{esc}}$ is directly proportional to the diameter of the planet. 
    The escape velocities from the billion year survivors shown here range from 1.9-2.7 km s$^{-1}$.
    The uncertainty is asymmetric.  The isothermal approximation predicts that a Europa with
    $v_{\mathrm{esc}}=2.2$ km s$^{-1}$ would retain its ocean for a billion years. 
    The coldest case has $v_{\mathrm{esc}}=1.9$ km s$^{-1}$, while the hottest case shown ($n=-0.04$) has
    $v_{\mathrm{esc}}=2.7$ km s$^{-1}$.  Still ``hotter'' cases are possible if photochemistry significantly
    reduces the molecular weight $m$.  
     The swath of uncertainty in $v_{\mathrm{esc}}$
     that envelopes the isothermal approximation appears to range between
      $-15\%$ to at least $+30\%$. 

  The right-hand panel of Figure \ref{three} moves the true Europa with respect to the Sun.    
    Here the billion year survivors are ice-covered Europas,
    at distances ranging between $0.85 - 1.4$ AU.  This corresponds to
    an uncertainty in insolation $I$ that is of the order of $\pm 60\%$.
    If photochemistry greatly reduces $m$, the evil $I$ could be smaller still. 
    We expect in reality that the swath of uncertainty in $I$ would be tighter than in Figure \ref{three},
    based on energy considerations. 
   Figure \ref{global_europa} above suggests that Europas with the warmest surfaces 
   will be more strongly influenced by condensation and will therefore
   have cooler temperature structures approaching the saturation curve, and thus will live longer 
   lives as marginally habitable planets than predicted by the isothermal
   approximation, whilst the colder Europas with thinner atmospheres
   will be more strongly influenced by solar EUV and FUV heating,
   and thus will be hotter and have shorter lifetimes than predicted by the isothermal approximation.
   
  \begin{figure}[!htb] 
   \centering
   \includegraphics[width=1.0\textwidth]{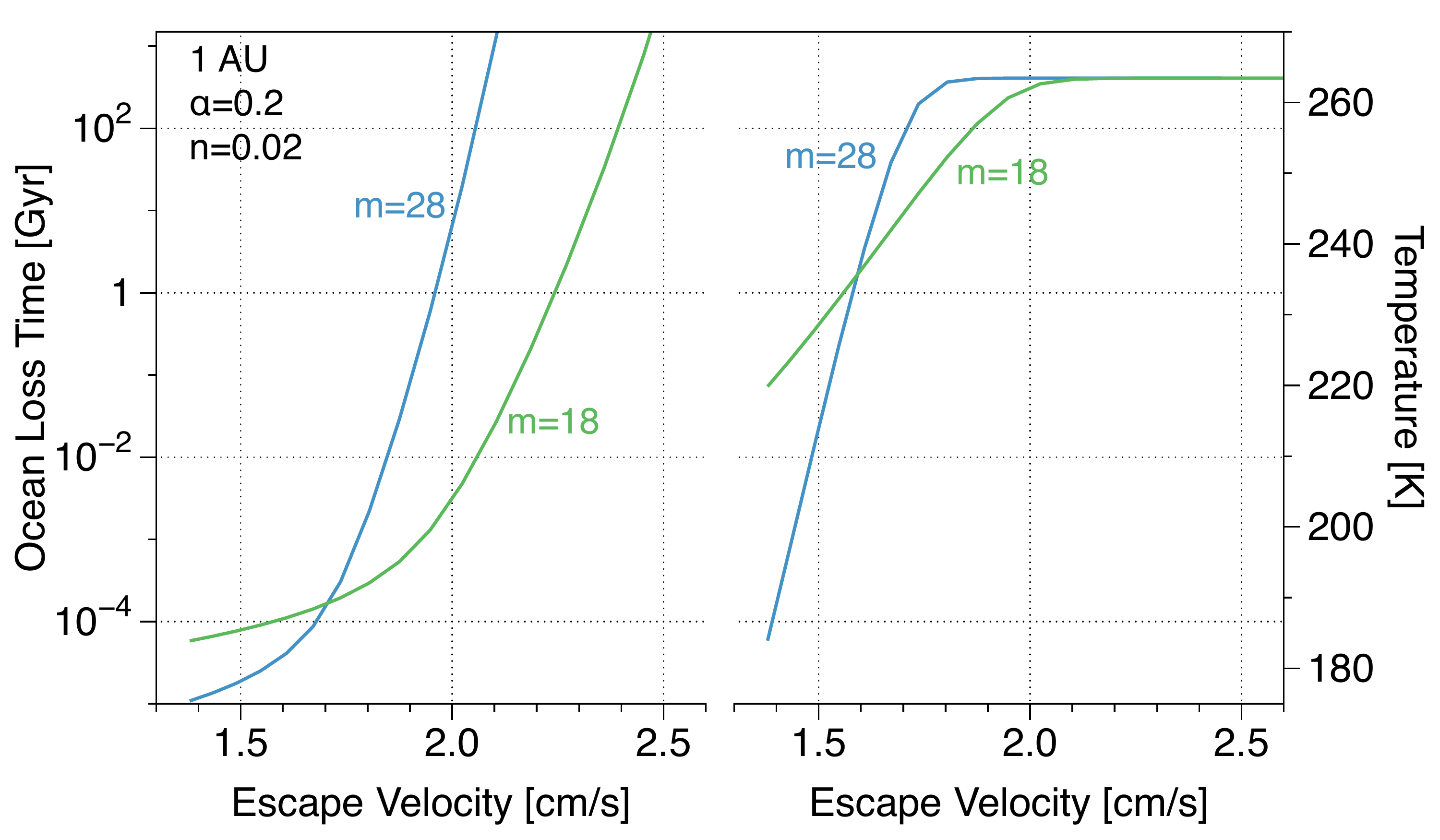} 
   \caption{\small The stabilizing effect of a ballast gas.
   Here we compare ``Europas'' of different sizes, one set with only water vapor ($m=18$)
   and the other with the addition of 1 bar of N$_2$ ($m=28$). 
    The left-hand panel shows ocean loss times,
    and the right-hand panel shows the surface temperatures.}
\label{ballast2}
\end{figure}

\subsection{Adding a heavy gas to increase the mean molecular weight}

It seems obvious that adding an abundant heavy constituent will reduce the scale height and reduce
the rate of escape.  For a tiny ocean world like Europa, the likeliest ballast gas is probably O$_2$,
which can be made directly from H$_2$O by selective hydrogen escape.
But the build up of O$_2$ is also determined atmospheric chemistry.  This is a matter
that we will address elsewhere.  For the present, we will simply consider a 1 bar N$_2$
atmosphere as a placeholder for examining the efficacy of a heavy gas as ballast.
With $p$N$_2 \gg p$H$_2$O, to first approximation the mean molecular weight is constant and
equal to nitrogen's 28. 
Water vapor $p$H$_2$O is assumed saturated at the surface.

For Figure \ref{ballast2} we set the albedo to 20\% and compare the pure water vapor
atmospheres $m=18$ to the N$_2$-dominated atmospheres $m=28$. 
Both models use $n=0.02$ polytropes.
Adding the heavy gas has an important stabilizing effect, corresponding
to an escape velocity that is 15\% smaller than if N$_2$ were not present.
The swath of uncertainty resulting from the presence or absence of other gases is comparable
to the uncertainty resulting from the use of the isothermal approximation.

\subsection{Discussion}
   The uncertainty in evaporation lifetimes of small icy worlds is asymmetric.
    It is unlikely that our coldest model is too cold.  That is, it is unlikely that the true 
    experience could be much more optimistic for continuous habitability than our most long-lived models. 
    But it is entirely possible that our hottest model falls well short of the true rate of escape,
    especially in the presence of a more active Sun.
    This is only partly because of temperature.
    The bigger uncertainty is photochemistry, which can reduce the mean molecular mass $m$ by
    a factor of three or more, to the detriment of alt-Europa as an abode of life.

\newpage


 \end{document}